\documentclass[manuscript]{aastex6}

\usepackage{amssymb}
\usepackage{amsmath}
\usepackage{xspace}

\begin{document}

\title{Formation of complex molecules in prestellar cores: a multilayer approach}

\author{A.I. Vasyunin}
\affil{Max-Planck-Institute for Extraterrestrial Physics, Garching, Germany}\affil{Ural Federal University, Ekaterinburg, Russia}\email{anton.vasyunin@gmail.com}
\author{P. Caselli}
\affil{Max-Planck-Institute for Extraterrestrial Physics, Garching, Germany}
\author{F. Dulieu}
\affil{LERMA, Universit\'e de Cergy Pontoise, Sorbonne Universit\'es, UPMC Univ. Paris 6, PSL Research University, Observatoire de Paris, UMR 8112 CNRS,  5 mail Gay Lussac 95000 Cergy Pontoise, France}
\author{I. Jim\'enez-Serra}
\affil{Queen Mary University, London, UK}

\begin{abstract}
We present the results of chemical modeling of complex organic molecules (COMs) under conditions typical for prestellar cores. We utilize an advanced gas-grain astrochemical model with updated gas-phase chemistry, with a multilayer approach to ice-surface chemistry and an up-to-date treatment of reactive desorption based on recent experiments of~\citet[][]{Minissale_ea16}. With the chemical model, radial profiles of molecules including COMs are calculated for the case of the prototypical prestellar core L1544 at the timescales when the modeled depletion factor of CO becomes equal to that observed. We find that COMs can be formed efficiently in L1544 up to the fractional abundances of 10(-10) wrt. total hydrogen nuclei. Abundances of many COMs such as CH$_{3}$OCH$_{3}$, HCOOCH$_{3}$, and others peak at similar radial distances of ~2000--4000 AU. Gas-phase abundances of COMs depend on the efficiency of reactive desorption, which in turn depends on the composition of the outer monolayers of icy mantles. In prestellar cores, the outer monolayers of mantles likely include large fractions of CO and its hydrogenation products, which may increase the efficiency of reactive desorption according to \citet[][]{Minissale_ea16}, and makes the formation of COMs efficient under conditions typical for prestellar cores, although this assumption is yet to be confirmed experimentally. The hydroxyl radical (OH) appears to play an important role in gas-phase chemistry of COMs, which makes it deserving further detailed studies.
\end{abstract}


\section{Introduction}
Evolution of organic matter and build-up of molecular complexity during the process of star- and planet formation is one of the most important but still not well understood questions in astochemistry \citep[][]{HerbstvanDishoeck09}. While the formation of exotic ``carbon-chain'' molecules known in cold interstellar cores for several decades can be explained via gas-phase ion-molecular and neutral-neutral reactions, the mechanisms of formation of saturated complex organic molecules (COMs) such as CH$_{3}$OH, CH$_{3}$OCH$_{3}$, HCOOCH$_{3}$ and others are more uncertain. Formation of COMs in hot cores and corinos is reasonably well explained by the ``warm-up scenario'' \citep[][]{GarrodHerbst06, Garrod_ea08}, in which complex organic molecules first form via surface radical-radical reactions at 30--40~K and then evaporate to the gas at higher temperatures reaching fractional abundances of about $\sim$10$^{-8}$--10$^{-7}$ with respect to hydrogen similar to those observed~\citep[][]{Blake_ea87, Cazaux_ea03, Bottinelli_ea04a, Bottinelli_ea04b}. However, organic molecules in cold cores \citep[e.g.,][]{Oeberg_ea10, Bacmann_ea12, Cernicharo_ea12, Vastel_ea14} require other mechanisms to explain their formation, since warm-up develops in later stages, after the switch on of a protostar.

To date, several scenarios have been proposed to explain the formation of COMs typical for terrestrial chemistry during the earliest stages of star formation prior to the warm-up phase. Note that all of them involve grain-surface chemical processes and non-thermal desorption of species from cold dust grains into the gas phase. The first attempt to explain terrestrial-type COMs discovered in L1689b and B1-b was made by \citet[][]{VasyuninHerbst13_coms}. In that study, the authors proposed that COMs in the cold gas may be formed via ion-molecule and radiative association reactions between precursor molecules formed on cold grains and then ejected in the gas phase via efficient reactive desorption. While the observed abundances of COMs were satisfactorily explained, the model significantly overestimated abundances of CH$_{3}$OH and H$_{2}$CO in the gas phase. \citet[][]{Balucani_ea15} improved this scenario by adding a neutral-neutral reaction linking methyl formate and dimethyl ether in the gas phase, and adjusting rates of several other important gas-phase reactions. These improvements allowed \citet[][]{Balucani_ea15} to reach better agreement with observations for methanol and establish a clear chemical link between CH$_{3}$OCH$_{3}$ and HCOOCH$_{3}$. \citet[][]{Reboussin_ea14} showed that impulsive heating of interstellar grains via cosmic ray particles may increase the mobility of species on grain surface and enhance formation of COMs, making their abundances somewhat closer to observed values. However, \citet[][]{Reboussin_ea14} do not take into account the locality of cosmic ray heating of a dust grain and icy mantle which has been shown to have major effect on the chemistry in the ice~\citep[][]{Ivlev_ea15}. Recently, \citet[][]{Ruaud_ea15} showed that the Eley-Rideal surface reaction mechanism which does not require surface mobility of species, and normally not considered in astrochemical models, assisted by complex induced reactions may be efficient in producing observed amounts of complex organic molecules at 10~K. However, the abundances of some important complex organic molecules such as methyl formate are below observed values by 1--2 orders of magnitude. Using microscopic Monte-Carlo simulations, \citet[][]{ChangHerbst16} showed that non-diffusive chain chemical reactions may lead to the formation of complex organic molecules in icy mantles of interstellar grains at low temperatures of $\sim$10~K. However, we note that the ejection of COMs from the grain surface to the gas phase does not consider the results from recent laboratory experiments (see below, and \citet[][]{Dulieu_ea13} and \citet[][]{Minissale_ea16b}).

Although models published to date have shed some light on the mystery of low-temperature formation of complex organic molecules, they rely on poorly known assumptions and parameters. At low temperatures of about 10~K, the only way to establish the required feedback from grain-surface chemistry to the gas phase chemistry is to invoke non-thermal desorption of species. Under the conditions typical for cold dark clouds, the most efficient type of non-thermal desorption is likely the so-called reactive desorption (a.k.a. chemical desorption, or desorption upon formation;~\citealt[][]{Garrod_ea06, Garrod_ea07}). By reactive desorption (RD) we call the process of breaking the surface-molecular bond of a reaction product due to the release of formation energy in surface two-body association reaction. Reactive desorption is a complex process which is controlled by a number of factors, e.g. internal structure of molecules, type of surface etc. The efficiency of reactive desorption (fraction of products of a surface reaction ejected into the gas phase) is a matter of debate in the community and a subject of both experimental and theoretical studies. In the early theoretical study by \citet[][]{Garrod_ea07} based on Rice-Ramsperger-Kessel (RRK) theory, the authors proposed the efficiency of RD to be of about 1--3\% and only slightly dependent on the type of desorbing molecule. \citet[][]{VasyuninHerbst13_coms} showed that for the explanation of observed abundances of COMs in cold clouds one needs to assume efficiency of RD of about 10\% at least for methanol and formaldehyde.

In a series of laboratory experiments on reactive desorption, which are the most comprehensive to date, \citet[][]{Dulieu_ea13} and \citet[][]{MinissaleDulieu14} found surprisingly high efficiency of reactive desorption for certain systems (e.g., 80\% for the reaction O+O$\rightarrow$O$_{2}$ on bare silicate) but also strong dependence of desorption efficiency on the particular chemical reaction and the type of surface. One of the important results for astrochemistry is that the efficiency of reactive desorption in laboratory experiments is dramatically reduced if bare silicate is covered by water ice. As summarized in \citet[][]{Minissale_ea16}, for the majority of studied reactions, the efficiency of reactive desorption falls below the upper limit of 5--10\% measurable in experiments, if surface is amorphous solid water. Using the experimental data, \citet[][]{Minissale_ea16} derived a semi-empirical theory that describes the dependence of the efficiency of RD on the type of surface, enthalpy of reaction and internal structure of desorbing molecule. The theory predicts negligible efficiency of reactive desorption for methanol and formaldehyde from the surface of water ice, which is believed to be the dominant constituent of icy mantles of grains in cold clouds \citep[e.g.,][]{Oeberg_ea11}. This fact can be considered as a serious argument against the scenario of formation of COMs in cold clouds proposed by \citet[][]{VasyuninHerbst13_coms}.

However, water ice is not the only major constituent of icy mantles in cold clouds. A number of observations over the last four decades revealed the main constituents of interstellar ices to be water~\citep[][]{GillettForrest73}, carbon monoxide~\citep[][]{Lacy_ea84}, carbon dioxide~\citep[][]{deGraauw_ea96}, methanol~\citep[][]{Grim_ea91}, ammonia~\citep[][]{vanDishoeck04} and methane~\citep[][]{Lacy_ea91}. The ice compositions appears to be somewhat different in protostellar objects at different stages of development, but surprisingly similar between objects from the same evolutionary class~\citep[][]{Oeberg_ea11}. In particular, the fraction of solid CO in ice is lower in high-mass protostars ($\sim$0.1), and significantly higher towards low-mass protostars, or prestellar cores ($\sim$0.3)~\cite[see Table~2 in][]{Oeberg_ea11}. From the analysis of the shapes of IR absorption bands, constraints on the ice structure were also inferred. In quiescent molecular dark clouds, e.g. in prestellar cores, icy mantles likely consist of two phases: water-rich polar phase and water-poor but CO-rich apolar phase~\cite[][]{Tielens_ea91}. It is likely that the CO-rich apolar phase is on top of water-rich mantles, because it is mainly formed from CO accreted from the dense gas after dense clouds are formed. Observed large ($\ge$90\%) depletion of gas-phase CO in prestellar cores~\citep[e.g.,][]{Caselli_ea99} and results of multilayer modeling of icy mantles~\citep[][]{GarrodPauly11, VasyuninHerbst13} are in favor of this assumption. As such, it is likely that the ice surface in prestellar clouds is mostly covered by CO ice, and, probably, by the product of CO hydrogenation such as formaldehyde and methanol. Indeed, \citet[][]{Bizzocchi_ea14} showed that the observed methanol towards the prestellar core L1544 exhibits low degree of deuteration compared to N$_{2}$H$^{+}$ and NH$_{3}$, which implies that its emission comes from the outer parts of the cloud where CO is on the onset of depletion. The CH$_{3}$OH map obtained by these authors toward L1544 shows a ring-like structure surrounding the dust continuum emission, which peaks at a distance similar to where CO depletion occurs within the core. As such, it is likely that CO accretion onto ices and the formation of CH$_{3}$OH and other COMs happens simultaneously, dominating the outer shells of icy mantles. This is in line with the recent results of \citet[][]{JimenezSerra_ea16} who have shown that COMs are enhanced toward this CH$_{3}$OH-rich, ring-like structure in L1544 with respect to its center. Efficiencies of RD from CO ice have not been studied experimentally so far. However, according to the semi-empirical theory by \citet[][]{Minissale_ea16}, they must be significantly higher than in the case of water ice. This makes prestellar cores excellent laboratories where to test whether the formation of COMs in cold objects occurs via the scenario proposed by \citet[][]{VasyuninHerbst13_coms}.

In this work, we present an extended model of formation of complex organic molecules in prestellar cores. The model by \citet[][]{VasyuninHerbst13_coms} is improved by adding a state-of-the-art treatment of reactive desorption by \citet[][]{Minissale_ea16}. For the first time, we theoretically explore the impact of the composition of ice surface on the efficiency of reactive desorption and, thus, on the formation of complex organic molecules. To better treat the composition of the surface and interior (bulk) of thick icy mantles of interstellar grains in dark clouds, we employ a multilayer approach to grain-surface chemistry, and treat grain-surface and bulk chemistry separately. In addition, we consider the impact of new gas phase reactions recently proposed by \citet[][]{Shannon_ea13,Shannon_ea14} and \citet[][]{Balucani_ea15} on the formation of complex organic molecules. By combining our chemical model with the physical model of the prestellar core L1544 by \citet[][]{KetoCaselli10}, we predict radial abundance profiles for complex organic molecules across this prototypical object.

The paper is organized as follows. In Section~\ref{section:model}, we describe our chemical model and employ the physical model of the prestellar core L1544. In Section~\ref{section:results}, modeling results are presented. Section~\ref{section:discussion} is devoted to the comparison with available observational data and discussion. Finally, in Section~\ref{section:summary} the summary of the study is given.

\section{Chemical and physical models}\label{section:model}
\subsection{A three-phase code with bulk chemistry}
In cold dark clouds, the thickness of icy mantles on interstellar grains is significant, and reach up to several hundreds monolayers~\citep[e.g.,][]{Gibb_ea04, GarrodPauly11, VasyuninHerbst13}. The formation of such thick mantles occurs on a timescale of at least several hundred thousands years~\citep[][]{GarrodPauly11, VasyuninHerbst13} and happens in parallel with ongoing physical evolution of a prestellar core. As such, it is likely that the composition of ice surface and outer layers of icy mantles is different from its bulk composition. Since one of the main goals of this study is to investigate the impact of the surface composition of icy mantles on the formation of COMs in prestellar clouds, it is reasonable to employ a multilayer approach to ice chemistry. This allows us to study the composition of icy mantles surfaces explicitly, and discriminate between chemistry that occurs on the surface and in the bulk of ice. This approach is qualitatively similar to that described by \citep[e.g.][]{Taquet_ea12}, \citet[][]{Garrod13} and \citet[][]{Ruaud_ea16}.

For this study, we modified the chemical code utilized in \citet[][]{VasyuninHerbst13_coms}. The code is based on chemical rate equations for gas phase chemical reactions and modified rate equations~\citep[][]{Garrod08, Garrod_ea09} for chemical reactions in icy mantles of interstellar grains. The evolution of abundances of gas-phase species is governed by the equation:
\begin{equation}\label{gaseq}
\frac{dn_i^{gas}}{dt}=\sum_{jk}k_{jk}n_j^{gas}n_k^{gas}-n_i^{gas}\sum_{il}k_{il}n_l^{gas}-k_{acc}n_i^{gas}+R_i^{des},
\end{equation}
where $n_{i(j,k)}^{gas}$ is the abundance of the $i(j,k)$-th species in the gas phase, $k_{jk}$ and $k_{il}$ are the rate constants of gas-phase reactions, $k_{acc}$ is the rate constant for accretion of the $i$-th species to grain surface, and $R_i^{des}$ is the rate of desorption of the $i$-th species from grains to the gas.

The equations governing the chemistry on interstellar grains are constructed in a way to take into account the complex structure of icy mantles. The following assumptions are made. In the icy mantle, two chemically distinct phases can be picked out: surface and bulk. In both phases, chemical reactions may occur. Species can be transferred between surface and bulk when the total number of molecules in the icy mantle is changed due to accretion or desorption or, alternatively, due to thermal diffusion of molecules in the mantle. The difference between the surface and the bulk is twofold. First, desorption of species to the gas phase is only allowed from the surface, i.e. a bulk species must appear on the surface of icy mantles before desorption. Second, diffusion of species on the surface is much faster than inside the bulk. \citet[][]{Garrod13} proposed that diffusion energy of species in the bulk is twice higher than on the surface. In this work, we utilize that value. The following pair of equations describe the chemical evolution of the $i$-th species in the icy mantle of an interstellar grain under the above assumptions:
\begin{equation}\label{surfeq}
\frac{dn_i^{surf}}{dt}=\left(\frac{dn_i^{surf}}{dt}\right)^{chem}+\left(\frac{dn_i^{surf}}{dt}\right)^{tran}+\left(\frac{dn_i^{bulk}}{dt}\right)^{diff} 
\end{equation}
\begin{equation}\label{bulkeq}
\frac{dn_i^{bulk}}{dt}=\left(\frac{dn_i^{bulk}}{dt}\right)^{chem}-\left(\frac{dn_i^{surf}}{dt}\right)^{tran}-\left(\frac{dn_i^{bulk}}{dt}\right)^{diff}
\end{equation}
In equations (\ref{surfeq}) and (\ref{bulkeq}), the first term describes the evolution of the abundance of the $i$-th species due to chemical reactions. In equation (\ref{surfeq}) the first term also includes processes of accretion and desorption. This term can be expressed either as a rate equation similar to that constructed for a gas-phase species, or as a modified version of a rate equation that is capable to take into account stochastic effects in surface chemistry \citep[][]{Garrod08}. Although the developed code is capable of using both types of equations, below we use modified rate equations for surface chemistry, unless otherwise stated. As shown in~\citet[][]{Garrod_ea09}, in contrast to standard rate equations, modified rate equations produce numerical results close to those obtained with the rigorous Monte Carlo approach even in the case of accretion-limited regime of surface chemistry at which stochastic effects are important. This regime is likely to be in action in this study due to the adopted parameters of surface chemistry (tunneling for H and H$_{2}$, see below).

The second term in equations (\ref{surfeq}) and (\ref{bulkeq}) is the rate of transfer of species between the surface and the bulk due to processes of accretion and desorption or change in number of molecules on surface due to chemical reactions, i.e. due to the processes that deposit or physically remove species to/from the icy mantle. In other words, this term represents the instant change of what is surface and what is bulk of the ice rather than the physical transfer of the material in the icy mantle. This term is defined in a way similar to that in previously developed models of e.g. \cite{HasegawaHerbst93} and \cite{GarrodPauly11}:
\begin{equation}
\left(\frac{dn_i^{surf}}{dt}\right)^{tran}=\frac{n_i}{N^{surf}}\cdot\left(\frac{dN^{surf}}{dt}\right)^{chem}\cdot\alpha_{tran}
\end{equation}
Here, $N^{surf}=\sum_i n_i$ is the total number of particles of species on the ice surface,
\begin{equation}
\left(\frac{dN^{surf}}{dt}\right)^{chem}=\sum_i\left(\frac{dn_i^{surf}}{dt}\right)^{chem}
\end{equation}
is the net rate of change of the total number of particles on the surface due to chemical reactions, accretion and desorption. Note that by varying $N^{surf}$ we can change the number of upper monolayers of the ice that belong to virtual ``surface'' in our model. This could be necessary to better satisfy experimental data, e.g. on desorption. Following finding of \citet[][]{VasyuninHerbst13}, we set $N^{surf}$ equal to four times the number of surface binding sites, thus assigning the upper four monolayers to the ``surface'', or the most chemically active fraction of ice.

The last term in equations (\ref{surfeq}) and (\ref{bulkeq}) describes real transfer of species between bulk and surface due to diffusion. In a thick mantle consisting of several hundreds of monolayers, an average atom or molecule must perform a series of jumps before reaching the surface due to 3D brownian motion. On each jump, a species has a chance to react with another reactive species or dissociate due to impact with a cosmic ray or photon absorption. As such, the surface is only reached by those species which do not undergo any chemical transformation along its way to the surface.

\citet[][]{Garrod13} defines the rate of diffusion of a species from bulk to surface with his Eq.~(3). Essentially, his definition assumes that any atom or molecule can reach the surface of the ice in one jump, regardless its initial position in the mantle. We believe that it is a too optimistic estimation, since it assumes that in a series of intramantle swaps needed for a species to reach the surface, no encounter with reactive species that would lead to a chemical reaction that will eliminate a swapping species will happen. Thus, the probability that the species will reach the ice surface rather than undergo a chemical reaction inside the bulk is $(1-P)^{N}$, where $P$ is the probability for a species to react on a single swap and $N$ is the number of swaps needed to reach the surface. Observational data on ice composition \citep[e.g.,][]{Oeberg_ea11} imply that the fraction of chemically active species such as CO that can react with the most mobile species (e.g., atomic hydrogen) is tenths of percents. Assuming efficient tunneling through activation barriers of surface reactions, including that with CO~\citep[][]{Hasegawa_ea92}, one can roughly assume that the swapping species can undergo a chemical reaction on each swap with a probability of 25\%, based on the average fraction of reactive species (mainly CO and H$_{2}$CO) obtained from ice observations~\citep[][]{Oeberg_ea11}; thus we get the probability of a species to reach the surface from the layers beneath the 10th below 5\%, e.g. $(1-0.25)^{11}=0.042$. Thus, we assume that species that start to migrate from deeper positions have a negligible probability to reach the surface when compared to the probability
to undergo intra-bulk chemical reactions. As such, the diffusion rate from bulk to surface (b2s) is defined as:
\begin{equation}
R_{diff,b2s}=n_{i}^{bulk}\cdot\frac{N^{surf}}{N^{bulk}}\cdot k^{swap}_i\cdot max\left( 1, \frac{10}{N_{ml}} \right)
\end{equation}
where $N^{bulk}$ is a total number of molecules in the bulk of ice, and $k^{swap}_i$ is the rate of swapping of a species in the bulk of ice, defined in the same way as in \citet[][]{Garrod13}, i.e., twice the surface diffusion rate of a species, $N_{ml}$ is the current number of monolayers in the mantle.

Photochemistry and chemistry induced by direct cosmic ray impacts are treated in the same way both on surface and in the bulk of ice, and is cloned from gas-phase processes included in the model. Although it is clearly a simplification of real physical processes~\citep[see e.g.,][]{Fayolle_ea11, MunozCaro_ea14, Ivlev_ea15}, we prefer this basic approach due to the lack of experimental data on details of surface photochemistry. Five types of desorption processes are included in the model: thermal desorption, cosmic ray--induced desorption~\citep[][]{HasegawaHerbst93}, cosmic ray-induced photodesorption~\citep[][]{PrasadTarafdar83}, photodesorption, and reactive (or chemical) desorption. Note that the efficiency of photodesorption in releasing the methanol from icy mantles of grains to the gas phase in the outer shells of prestellar cores, is quite uncertain. Although experiments by~\citet[][]{Oeberg_ea09} showed high efficiency of photodesorption with a yield of 10$^{-3}$ molecule per incident photon, newer experiments by \citet[][]{Bertin_ea16} and \citet[][]{CruzDiaz_ea16} suggest that at least for methanol the yield does not exceed 10$^{-5}$ molecule per incident photon, thus making the photodesorption process less efficient than e.g. ice sputtering by cosmic rays, as shown by~\citet[][]{Dartois_ea15}. Given the discrepancy between the old and new experimental results, in the current study we assumed a photodesorption yield equal to 10$^{-5}$ molecule per incident photon.

\subsection{Surface formation of methanol and formaldehyde}\label{section:methanol_formation}
The network of surface reactions used in this work is taken from \citet[][]{VasyuninHerbst13_coms}. In this network, formaldehyde and methanol are formed on surface via the hydrogenation of CO molecule in the sequence CO~$\rightarrow$~HCO~$\rightarrow$~H$_{2}$CO~$\rightarrow$~CH$_{2}$OH~$\rightarrow$~CH$_{3}$OH,
which was experimentally confirmed by e.g. \citet[][]{Fuchs_ea09}. Hydrogen addition reactions leading to the formation of HCO and CH$_{2}$OH have activation barriers of 2500~K~\citep[][]{RuffleHerbst01} (see below Section~\ref{section:phys_cond} for the discussion on the shape of the reaction activation barriers). Note that this hydrogenation sequence of CO forming formaldehyde and methanol may not reflect the actual complexity of the CO hydrogenation process. Recently, \citet[][]{Minissale_ea16c} showed that hydrogenation of HCO and H$_{2}$CO is a more complex process accompanied by ``backward'' H$_{2}$ abstraction reactions H~+~HCO~$\rightarrow$~CO~+~H$_{2}$ and H~+~H$_{2}$CO~$\rightarrow$~HCO~+~H$_{2}$. These processes may increase the efficiency of the reactive desorption of CO and suppress the formation of methanol. Although these newly introduced processes may be of importance for the formation of COMs in cold clouds, they are not included in the present study. Branching ratios between H abstraction and H addition reactions have not been measured accurately yet \citet[][]{Minissale_ea16c}. Thus, the incorporation of abstraction reactions with poorly known rates will not reduce the resulting uncertainty of our model. Mechanism of surface formation of methanol is a problem of fundamental importance in astrochemistry. A separate study must be devoted to its reconsideration.

\subsection{Treatment of reactive desorption}
The treatment of desorption in our three-phase model is similar to that described in \citet[][]{VasyuninHerbst13_coms} with the exception of reactive desorption. For this type of desorption, we adopt the recent results by~\citet[][]{Minissale_ea16}. They derived a semi-empirical formula that describes the dependence of the efficiency of reactive desorption on the surface composition, exothermicity of a surface reaction and binding energies of reaction products:
\begin{equation}\label{rdeff}
R_{RD} =\exp\left(-\epsilon\frac{E_{b}\times DF}{\Delta H}\right), \epsilon=\left(\frac{m-M}{m+M}\right)^2.
\end{equation}
Here, ${\rm R}_{RD}$ is the efficiency of reactive desorption, i.e., the fraction of products of a surface reaction directly ejected to the gas, ${\rm E}_{b}$ is the binding energy of a species, $DF$ is the number of vibrational modes in a molecule--surface bond system, $\epsilon$ is the fraction of kinetic energy retained by the reaction product with the mass $m$ colliding with the surface element with effective mass $M$. Experiments on scattering of molecules on surface~\citep[][]{Hayes_ea12} showed that upon collision, molecules interact with a surface structural element consisting of several atoms or molecules forming the surface, not with a single one. This is due to collective effects caused by the surface rigidity. As such, the effective mass of the surface element $M$ is typically much higher than the mass of single atoms or molecules the surface consists of.

The formula (\ref{rdeff}) has been established for rigid surface (graphite), where an effective mass has been measured by other experimental methods to be close to 1.8 masses of graphite carbon ring (M$\approx$130 a.m.u.; \citet[see discussion in][]{MinissaleDulieu14}. Application of formula~(\ref{rdeff}) for molecular surfaces also requires assigning a collective mass larger than the individual molecules constituent of the ice, because, e.g., chemical desorption of H$_{2}$O is still measurable from H$_{2}$O surface, whereas the strict use of the formula gives zero~\citep[][]{Minissale_ea16}. Given the much lower efficiencies of reactive desorption from water surface observed in experiments, the value of M for water surface has been chosen equal to 48 a.m.u. as best matching data from experiments~\citep[][]{Minissale_ea16}. Recent experiments involving surfaces made of CO and H$_{2}$CO ice, showed relatively efficient reactive desorption of these two species into the gas~\citep[][]{Minissale_ea16c}, thus advocating quite high effective mass for such surfaces. Any equivalent mass between 80 and 120, indeed agrees with these measurements. Also, in case of CO ice, the binding energy of adsorbates may be lowered with respect to water ice, which also contributes to raise the chemical desorption efficiency on CO ice. Thus, in order not to include any new poorly controlled parameter in the model, we have chosen an effective mass M for CO ice equal to 100. Although variations of M in the range 80--120 may change the maximum values of modeled abundances within an order of magnitude, they will not change the general conclusions of this work.

\begin{figure}\centering
  \includegraphics[width=0.60\textwidth, angle=90]{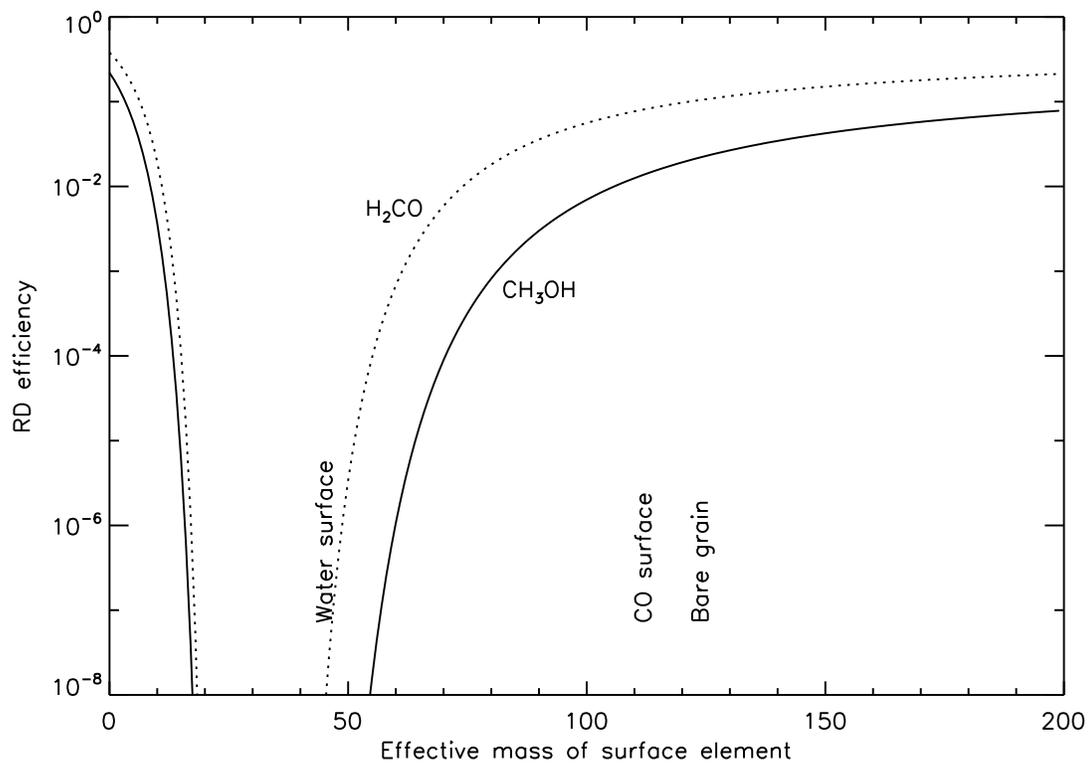}\vspace{5mm}
  \caption{Dependence of the efficiency of reactive desorption of methanol and formaldehyde in reactions H~+~HCO~$\rightarrow$~H$_{2}$CO and H~+~H$_{3}$CO~$\rightarrow$~CH$_{3}$OH on the chosen effective mass of the surface element (based on \citet[][]{Minissale_ea16b}; see text for details)}.
  \label{fgr:rdeff}
\end{figure}

The fraction of kinetic energy retained by the reaction product $\epsilon$ is an important parameter that governs the dependence of the efficiency of reactive desorption on surface composition. Its physical meaning is the fraction of kinetic energy retained by a molecule with mass $m$ upon collision with a structural element of the surface with mass $M$, under the approximation of classic elastic collision. As such, the efficiency of reactive desorption depends not only on the type of desorbing molecule and enthalpy of reaction, but also on the surface composition, e.g., on the outer layers composition of a thick icy mantle. In Figure~\ref{fgr:rdeff}, the efficiency of the reactive desorption is shown as a function of M for two key species, H$_{2}$CO and CH$_{3}$OH. The vertical text, ``CO ice'', ``water ice'' and ``bare grain'' indicate the $M$ value for the corresponding surface types. We assume that the value for ``CO ice'' is also used for species of similar molar weight, such as N$_{2}$, H$_{2}$CO and CH$_{3}$OH, since these species have similar molecular masses and no experimental data on scattering of molecules on surfaces consisting of those species are available. These species are expected to constitute a significant fraction of an ice surface under the conditions typical of prestellar cores. Therefore, the total efficiency of reactive desorption is calculated as a sum of efficiencies of reactive desorption on different types of surfaces: bare grain, water ice, ``heavy ice'' (CO + N$_{2}$ + H$_{2}$CO + CH$_{3}$OH ices) times the fraction of the corresponding surface type within the whole surface.

\subsection{Updates to the chemical network}
Since \citet[][]{VasyuninHerbst13_coms}, several important studies concerning chemistry of complex organic molecules in cold environments were published. We updated our chemical network with new gas-phase reactions proposed in \citet[][]{Shannon_ea13,Shannon_ea14} and \citet[][]{Balucani_ea15}.
Namely, \citet[][]{Shannon_ea13, Shannon_ea14} discovered that reactions between hydroxyl radical OH and oxygenated hydrocarbons such as methanol and dimethyl ether are efficient at low temperatures due to tunneling through activation barrier and formation of a hydrogen-bonded complex:
\begin{equation}
{\rm OH}+{\rm CH_{3}OH}\rightarrow {\rm CH_{3}O} +{\rm H_{2}O}
\end{equation}
\begin{equation}\label{oh_ch3och3}
{\rm OH}+{\rm CH_{3}OCH_{3}}\rightarrow {\rm CH_{3}OCH_{2}} +{\rm H_{2}O}
\end{equation}

Also, \citet[][]{Balucani_ea15} proposed a gas-phase chemical link between dimethyl ether (CH$_{3}$OCH$_{3}$) and methyl formate (HCOOCH$_{3}$) through the radical CH$_{3}$OCH$_{2}$:
\begin{equation}
({\rm F}, {\rm Cl})+{\rm CH_{3}OCH_{3}}\rightarrow {\rm CH_{3}OCH_{2}} +({\rm HF}, {\rm HCl})
\end{equation}
\begin{equation}
{\rm CH_{3}OCH_{2}}+{\rm O}\rightarrow {\rm HCOOCH_{3}}+{\rm H}
\end{equation}
The discovery of a new class of gas-phase reactions with hydroxyl radical efficient at low temperatures as well as a chemical link between dimethyl ether and formaldehyde are important findings for the problem of formation of COMs. To the best of our knowledge, this combination of new reactions has not been included into a full-scale astrochemical model before.

The gas-phase formation route of formamide (NH$_{2}$CHO) has been updated according to the recent studies by \citet[][]{Barone_ea15} and \citet[][]{Skouteris_ea17}. Using quantum chemical calculations, \citet[][]{Skouteris_ea17} updated the previous result by \citet[][]{Barone_ea15}, and estimated the rate constants for the reaction
\begin{equation}\label{nh2_h2co}
{\rm NH_{2}}+{\rm H_{2}CO}\rightarrow {\rm NH_{2}CHO}+{\rm H}
\end{equation}
to be equal to $\alpha$=7.8$\times$10$^{-16}$~cm$^{3}~s^{-1}$, $\beta$=-2.56, $\gamma$=4.88~K. This gives the value of modified Arrhenius rate constant which is defined as $k=\alpha\cdot(T/300K)^{\beta}\cdot\exp(-\gamma/T)$ equal to 2.9$\times$10$^{-11}$~cm$^{3}$s$^{-1}$ at 10~K. As such, reaction (\ref{nh2_h2co}) is expected to be a major route of formation of formamide in cold clouds (as long as NH$_{2}$ and H$_{2}$CO are abundant in the gas phase).

Finally, we altered the gas-phase chemistry of acetaldehyde (CH$_{3}$CHO) by including four gas-phase reactions in our model:
\begin{equation}\label{ch_ch3oh_1}
{\rm CH} + {\rm CH_{3}OH}\rightarrow {\rm CH_{3}CHO} + {\rm H},
\end{equation}
\begin{equation}\label{ch_ch3oh_2}
{\rm CH} + {\rm CH_{3}OH}\rightarrow {\rm CH_{3}} + {\rm H_{2}CO},
\end{equation}
\begin{equation}\label{c_ch3cho}
{\rm C} + {\rm CH_{3}CHO}\rightarrow {\rm C_{2}H_{4}} + {\rm CO}
\end{equation}
and
\begin{equation}\label{ch3_hco}
{\rm CH_{3}}+{\rm HCO}\rightarrow {\rm CH_{3}CHO} +h\nu
\end{equation}
Processes (\ref{ch_ch3oh_1}) and (\ref{ch_ch3oh_2}) are the two most likely outcomes of the reaction between CH and CH$_{3}$OH experimentally studied by \citet[][]{Johnson_ea00}. They found this reaction to be barrierless, pressure-independent with negative dependence on temperature. The total rate of the reaction between CH and CH$_{3}$OH at 300~K is reported by \citet[][]{Johnson_ea00} to be  2.5$\times$10$^{-10}$~cm$^{3}s^{-1}$ with a temperature dependent Arrhenius factor -1.93. Since channel (\ref{ch_ch3oh_2}) is more thermodynamically favorable than channel (\ref{ch_ch3oh_1}), we assume 90\% probability for the channel (\ref{ch_ch3oh_2}) to occur vs. 10\% probability for the channel (\ref{ch_ch3oh_1}) (see Table~\ref{tbl:newr}). Interestingly, channel (\ref{ch_ch3oh_1}) is added to the UDFA12 database~\citep[][]{McElroy_ea13}, but is absent in KIDA database~\citep[][]{Wakelam_ea15}. On the contrary, channel (\ref{ch_ch3oh_2}) does not exist in UDFA12, but added to KIDA.

Reaction (\ref{c_ch3cho}) is taken from \citet[][]{HusainIoannou91}. This reaction is expected to reduce the gas-phase abundance of acetaldehyde at early times when atomic carbon is relatively abundant in the gas phase. A similar reaction between carbon atoms and formaldehyde from \citet[][]{HusainIoannou91} is also added to the chemical network to better constrain the abundance of H$_{2}$CO at early times.

The rate for reaction (\ref{ch3_hco}) has been taken from \citet[][]{CallearCooper90}, who studied the reaction experimentally, under the temperature range 373~K~---~473~K. We assume that the rate of reaction (\ref{ch3_hco}) remains the same down to 10~K: 5$\times$10$^{-11}$~cm$^{3}$~s$^{-1}$, although, it is possible that at low temperature the rate of the reactions could be higher due to more efficient stabilization of initially vibrationally excited CH$_{3}$CHO. The details of all newly included reactions are summarized in Table~\ref{tbl:newr}.

\subsection{Physical model of L1544 and initial conditions for chemistry}\label{section:phys_cond}
\begin{figure}\centering
  \includegraphics[height=0.80\textwidth, angle=90]{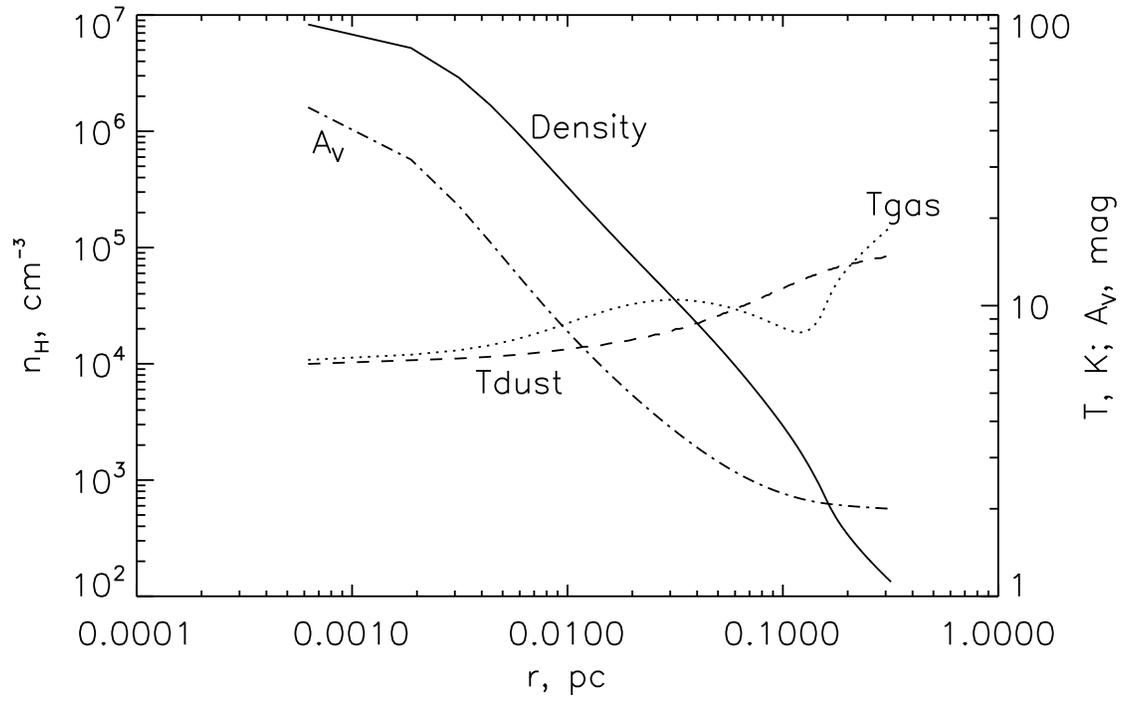}\vspace{5mm}
  \caption{Physical structure of L1544 from \citet[][]{KetoCaselli10}. Note that gas and dust temperature slightly decrease from the edge to center from $\approx$20~K to $\approx$~7~K, while the visual extinction rises sharply from 2 mag to more than 50 mag.}
  \label{fgr:L1544_structure}
\end{figure}

To model the radial profiles of COMs in L1544, we use the results from a 1D physical model of the prestellar core developed by \citet[][]{KetoCaselli10}, see Figure~\ref{fgr:L1544_structure}. In the core, one can single out the very dense and dark central part, an intermediate shell with moderate density, and the outer part with low density and penetrating UV field. We took the visual extinction at the edge of the core equal to A$_{V}$=2~mag to simulate the fact that L1544 is embedded in a molecular cloud. In total, our physical model has 128 radial points with different physical conditions. For each point, chemistry has been calculated independently using our 0D chemical model. Single grain size of 10$^{-5}$~cm is used to calculate visual extinction and chemical evolution in the current study.

To obtain the initial abundances of the molecular species at t=0 yr in our simulations, we ran the chemical code for the time span of 10$^{6}$~years using physical conditions typical for a translucent cloud: hydrogen column density n$_{H}$=10$^{2}$~cm$^{-3}$, temperature T=20~K, visual extinction A$_{V}$=2~mag, and ``low metals'' atomic initial conditions, corresponding to the values listed as EA1 in Table~1 in \citet[][]{WakelamHerbst08}. Final abundances from these simulations were taken as initial values for the simulations presented in this study.

The mobility of species on grain surfaces is a matter of debate in the astrochemical community. At low temperatures typical for prestellar clouds, only a limited number of species should be mobile on grain surfaces. Among them, atomic and molecular hydrogen. For these species, the main source of mobility is assumed to be quantum tunneling through a rectangular barrier of width 1\AA~separating the binding sites~\citep[e.g.,][]{Hasegawa_ea92}. Although the efficiency of quantum tunneling for H and H$_{2}$ on grain surfaces has been debated for a long time, no solid agreement is reached on this point so far~\citep[see e.g., for review,][]{HamaWatanabe13}. In this study, we enabled tunneling for diffusion of H and H$_{2}$ in the model. The importance of tunneling for the simulations results is highlighted in Section~\ref{section:discussion}. Tunneling was also claimed to be efficient for oxygen atoms at low temperatures~\citep[][]{Minissale_ea13}. However, in a more recent experiment by \citet[][]{He_ea14}, mobility of atomic oxygen was only observed at T$\sim$40~K, which contradicts the assumption of efficient tunneling for O atoms. As such, we have chosen not to include atomic oxygen tunneling in our model. For other species, the main diffusion mechanism is due to thermal hopping, with a rate defined by the binding--desorption energy ratio, E$_{b}$/E$_{D}$. Following recent estimates by~\citet[][]{Minissale_ea16b}, we select the value of E$_{b}$/E$_{D}$=0.55.
However, at dust temperatures $\sim$10~K with enabled tunneling for H and H$_{2}$ diffusion, this value is not expected to play a crucial role for the chemical evolution of icy mantles, even when set to the lowest value considered in literature, E$_{b}$/E$_{D}$=0.3~\citep[][]{Hasegawa_ea92}. As mentioned above, for bulk chemistry, this value is doubled. However, at low dust temperatures, bulk chemistry is well separated from surface chemistry, and all the essential processes affecting abundances of gas-phase species discussed below are impacted by the composition of the surface layers of icy mantles. This is explicitly simulated in the model in a framework of a three-phase formalism ``gas-surface-bulk''.

Tunneling through potential barriers is not only important for surface mobility of light species such as H and H$_{2}$, but also for calculations of the rates of reactions with activation barriers. Several approaches have been proposed to calculate the transmission probabilities through the reaction activation barriers on surface. \citet[][]{Hasegawa_ea92} proposed to calculate the transmission probability through the rectangular activation barrier of 1~\AA~~ width, while \citet[][]{GarrodPauly11} proposed to use the width of the rectangular activation barrier of 2~\AA. Also, \citet[][]{Taquet_ea13}, recently employed the Eckart's model to calculate the transmission probability. Here, we use transmission probabilities through rectangular barriers with widths of $a$~=~1.2~\AA. Our simulations show that the composition of the gas phase in the considered model is stable with respect to the choice of the width $a$ in the range [1.0~---~1.5]~\AA, while the composition of the bulk of the ice is quite sensitive to the exact value of $a$, and resemble the observed ice composition best at $a$~=~1.2~\AA. As such, by the choice of the particular value of $a$ we do not aim to put exact constraints on transmission probabilities for reactions with activation barriers in interstellar ice. Instead, we rather aim to show that our model is capable of reproducing simultaneously abundances of COMs in L1544, and produce a reasonable composition of the icy mantles of interstellar grains.

\section{Results}\label{section:results}
\subsection{Radial profiles of chemical abundances}
Observed values of fractional abundances of species towards prestellar cores and other astronomical objects are usually inferred from the observed column densities of species divided by the column density of H$_{2}$. This provides abundances averaged over the line of sight. As such, in compact objects with strong gradients of physical parameters such as prestellar cores, the abundances derived from column densities differ from true local fractional abundances. However, since the goal of this work is to investigate chemistry of complex organic molecules in different parts of L1544, below in most cases we present modeled true local fractional abundances of species calculated as volume concentration of a species divided by volume concentration of H$_{2}$ vs. time and radius. When presented, column densities are calculated following the expressions (1) and (2) in Section~7 in \citet[][]{JimenezSerra_ea16}.

Since abundances of species evolve with time in each point of L1544, it is important to determine the time which gives best agreement with observations. Our model does not include any treatment of dynamical evolution of L1544, neither treatment of collapse nor mixing between gas parcels. Also, the ``chemical age'' of an astrophysical object does not necessarily correspond to its physical age. Therefore, we have chosen the depletion factor of CO towards the center of L1544 as an estimation of the time for which the comparison of modeled chemistry with observations should be performed. The depletion factor of CO, defined as the ratio between the reference and observed CO column densities, was measured by \citet[][]{Caselli_ea99} as $\sim$10 toward the dust peak position, the center of our spherically symmetric core model. This depletion factor of CO can be easily calculated as a function of time in our model.  Figure~\ref{fgr:co2h2coldens} shows that the CO depletion factor becomes similar to the observed one after a time of $\sim$10$^{5}$~years. Therefore, we present modeled radial abundances of species in L1544 for this time, unless otherwise stated.

\begin{figure}
\includegraphics[height=0.80\textwidth, angle=90]{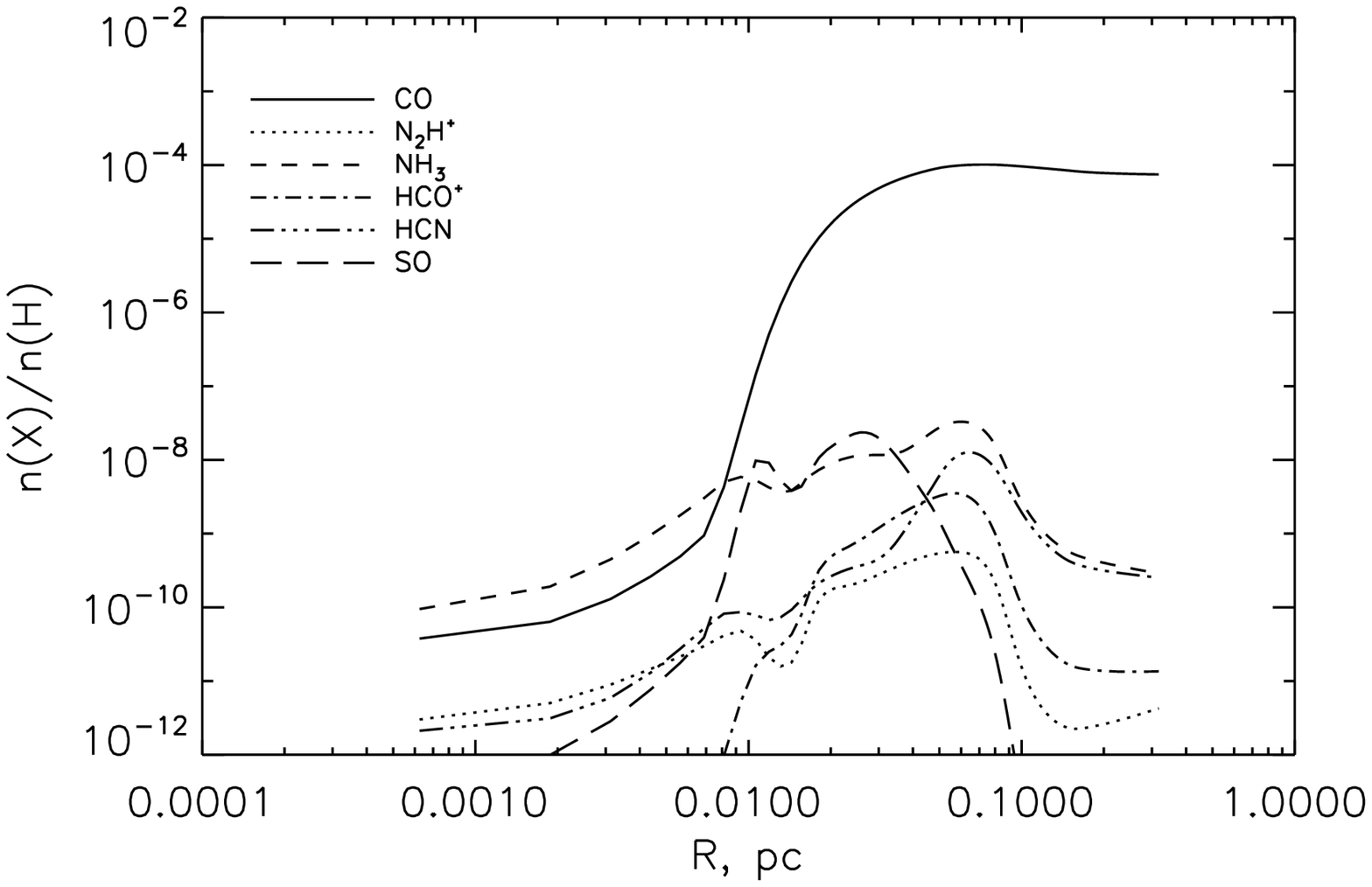}
\includegraphics[width=0.80\textwidth, angle=0]{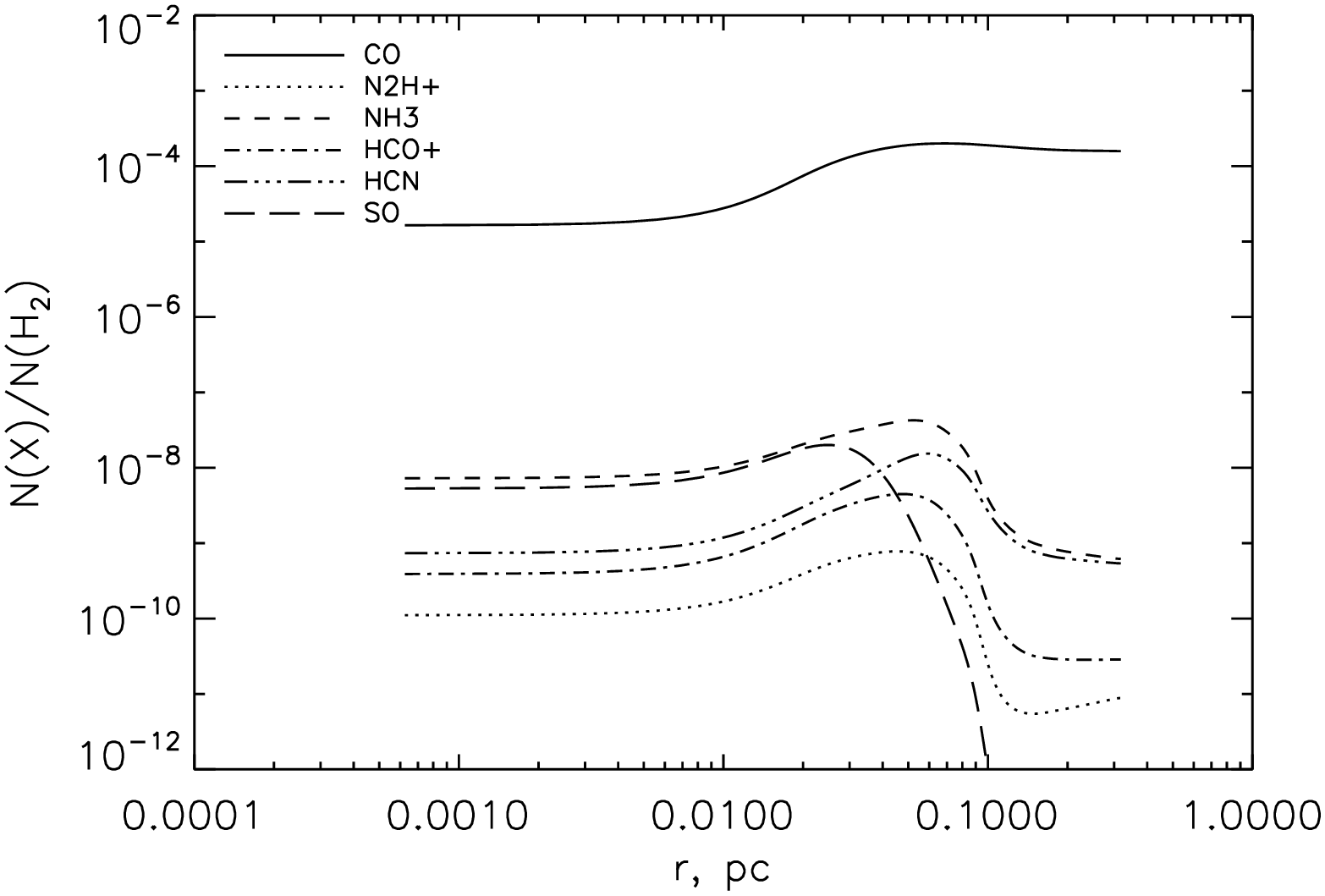}
\caption{Radial profiles of abundances (top panel) and convolved with a telescope beam size of 26'' column densities (bottom panel) of simple species after 10$^{5}$~years of chemical evolution in L1544.}\label{fgr:simple_species_vs_r}
\end{figure}

In Figure~\ref{fgr:simple_species_vs_r}, the top panel, radial profiles of selected simple species typically observed towards cold dark clouds are shown. By 10$^{5}$~years, all simple molecules except CO exhibit hump-like profiles. Towards the center of L1544, species are significantly depleted due to high density and low temperature. In the outer parts of the cloud, at radial distances R$\ge$0.03~pc, the visual extinction drops below 3~mag, thus making most molecules vulnerable to photodissociation by interstellar UV field. H$_{2}$ and CO are the exception, because of the effect of self-shielding and shielding by dust. The gas-phase abundance of CO drops sharply at R$\le$0.03~pc, where the gas density is higher than 10$^{5}$~cm$^{-3}$, in agreement with findings by \citet[][]{Caselli_ea99}. Abundance profiles of two key nitrogen-bearing species in prestellar clouds, NH$_{3}$ and N$_{2}$H$^{+}$, exhibit great similarity, with ammonia being about two orders of magnitude more abundant. This behaviour is typical of low-mass dense cores, including L1544, as found by~\citet[][]{Tafalla_ea02}. Radial behavior of HCO$^{+}$ is similar to the best fit Model~3 in \citet[][]{Caselli_ea02b}. N$_{2}$H$^{+}$ and HCO$^{+}$ are the two most abundant positive ions, thus reflecting the ionization degree. In turn, it appears similar to the measurements of \citet[][]{Caselli_ea02b}. In the bottom panel of  Figure~\ref{fgr:simple_species_vs_r}, radial profiles of column density ratios of simple species wrt. column density of H$_{2}$ convolved with a telescope beam size of 26'' are shown: N(X)/N(H$_{2}$). This corresponds to the abundances of species directly inferred from observations by \citet[][]{JimenezSerra_ea16}. One can see that column density profiles show flatter profiles than local fractional abundances of species. Indeed, the CO-to-H$_{2}$ column density ratio decreases by an order of magnitude toward the outer part of the core, instead of six orders of magnitude for local volume abundances. For NH$_{3}$, the N(NH$_{3}$)/N(H$_{2}$) ratio decreases by an order of magnitude instead of two and a half orders for local volume abundances. Such even behavior of column density ratios match observational data on molecular distributions in L1544 better than local volume abundances, especially for NH$_{3}$, for which observations are consistent with N(NH$_{3}$)/N(H$_{2}$) increase towards the center of the core \citep[][]{Tafalla_ea04}, while the N(N$_{2}$H$^{+}$)/N(H$_{2}$) column density ratio appears to be constant. This ``extra'' depletion of N-bearing species compared to observations may be due to the fact that dynamics is ignored in our simulations (see Section~\ref{section:discussion} for further discussion). Such behavior for N-bearing molecules was also present in our previous chemical models. Therefore, updates to our model concerning efficient reactive desorption and new gas-phase chemistry of COMs does not affect noticeably well-established chemistry of simple species in the gas phase, and we do not discuss it further.

\begin{figure}
  \includegraphics[height=0.8\textwidth, angle=90]{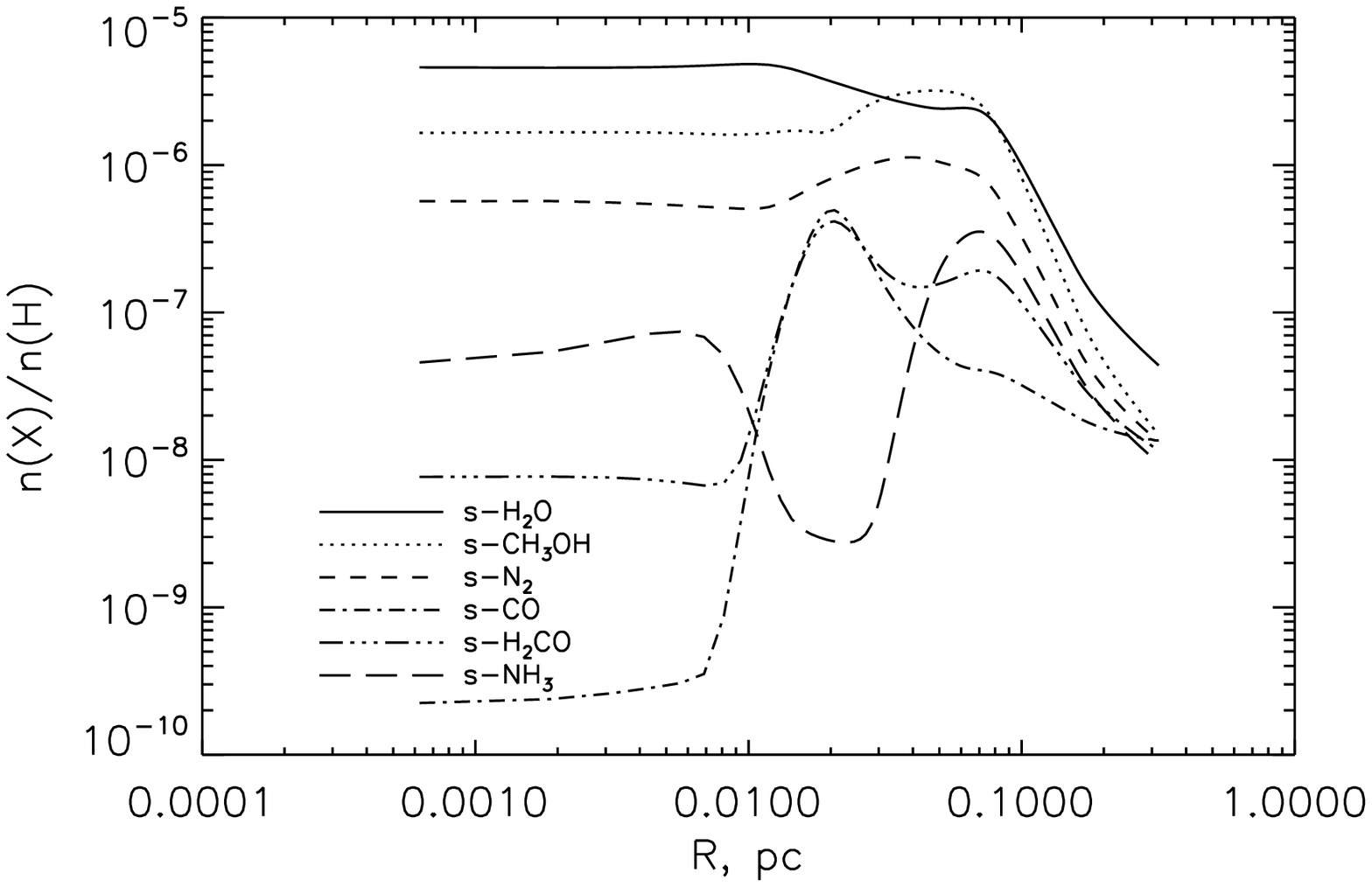}\newline
  \includegraphics[height=0.8\textwidth, angle=90]{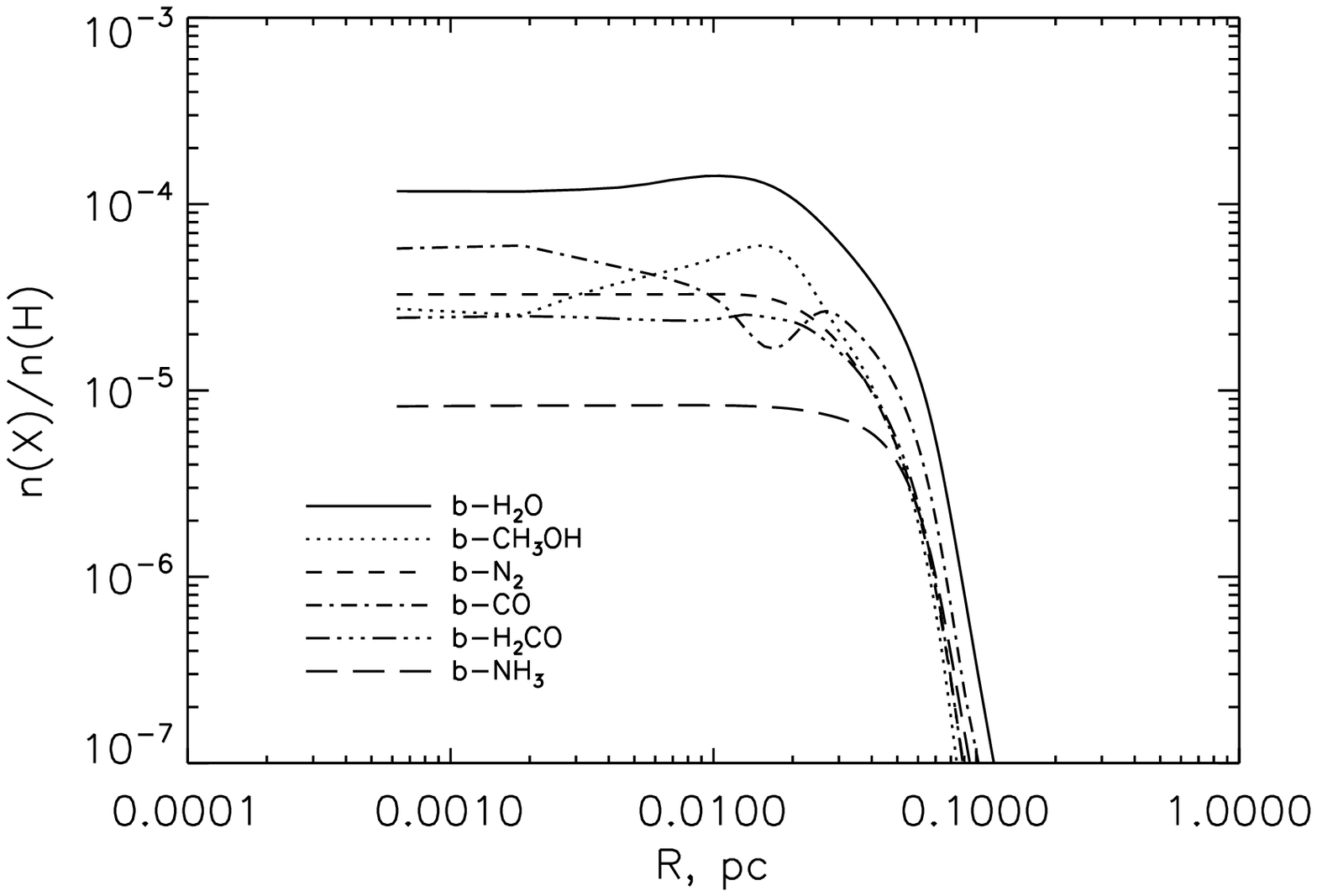}
  \caption{Composition of the surface ice layers (top) and bulk ice (bottom) at L1544 vs. radius at 1.6$\times$10$^{5}$ years.}
  \label{fgr:ice}
\end{figure}

Let us now consider the chemistry in icy mantles of interstellar grains. Advanced treatment of icy mantles utilized in this study, allows us to consider the chemical evolution of ice surface and bulk ice separately. Since mobility of species on the surface and inside the bulk is different, and desorption is only possible from the surface, bulk and surface usually have different chemical composition. In Figure~\ref{fgr:ice}, composition of ice surface (top panel) and bulk ice (bottom panel) is shown vs radius after 10$^{5}$~years of evolution. Plotted species are the most abundant ones on surface and in the bulk ice respectively, at radial distance of R=0.02~pc.

The ``surface'' of icy mantles in our model consists of a number of molecules corresponding to four monolayers of ice. We prefer not to distinguish individual monolayers in the ice, because bulk diffusion of species is enabled in the model. Under this assumption, individual monolayers cannot be unambiguously identified. Besides the new treatment of reactive desorption, we did not introduce any new chemical processes in our model. Therefore, the chemical evolution on ice surface is in general similar to that described in previous studies~\citep[][]{GarrodPauly11, VasyuninHerbst13}.

\begin{figure}
  \includegraphics[height=0.8\textwidth, angle=90]{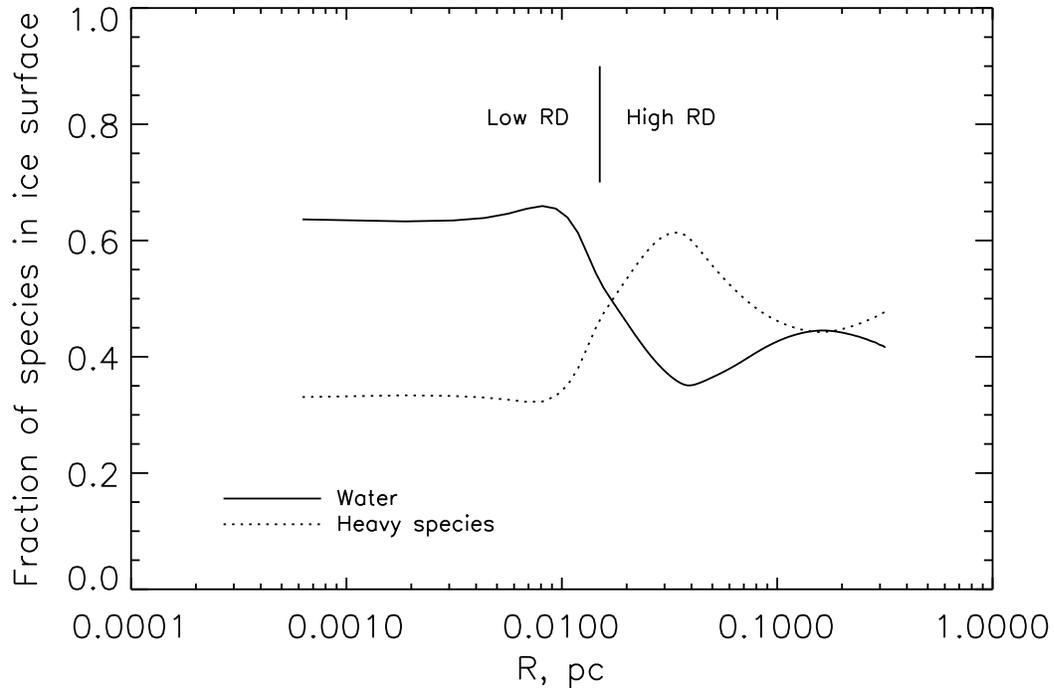}
  \caption{Fraction of water and ``heavy species'' in the surface layers of icy mantles of grains vs. radius after 10$^{5}$ years of evolution. Reactive desorption is only efficient from the fraction of surface consisting of ``heavy species''.}
  \label{fgr:heavy-light-ice}
\end{figure}

In the entire range of distances from the center of the cloud, the chemical composition of ice surface does not differ significantly.
Since in our model we enable quantum tunneling for H and H$_{2}$ and diffusion/desorption energy ratio E$_{b}$/E$_{D}$=0.55, the major constituents of the ice surface are water, methanol, molecular nitrogen as well as somewhat less abundant carbon monoxide, formaldehyde and ammonia. While CO and N$_{2}$ are mainly accreted from the gas phase, other major species are formed on surface in hydrogenation reactions. Water is mainly formed in surface reaction between atomic hydrogen and hydroxyl radical, but is also accreted from the gas phase but in less extent. Methanol and formaldehyde are formed during hydrogenation of CO molecule as described in Section~\ref{section:methanol_formation}. Quantum tunneling for diffusion and passing through activation barriers of reactions with H and H$_{2}$ makes CO hydrogenation efficient, thus leading to the abundance of CH$_{3}$OH on surface higher than abundances of CO and H$_{2}$CO. Note that, if tunneling for diffusion or passing through reaction activation barriers is neglected, our model cannot produce methanol in the ice in appreciable amounts. Interestingly, at timescales after 10$^{5}$~years, methanol and water on surface become chemically related, because significant fraction of atomic hydrogen and hydroxyl radical on surface at that time is produced by the photodissociation of methanol by cosmic ray-induced photons in the center of the core, or by interstellar UV photons at the edge of the cloud. The fraction of surface covered with ``heavy'' species (mainly methanol) vary from $\sim$0.35 at R$\le$0.01~pc to $\sim$0.65 at outer radii (Figure~\ref{fgr:heavy-light-ice}). This shift of balance between water and heavy species is mostly due to slightly different rates of photodissociation of water and methanol. In our model, the exponential factor for photorates $\gamma_{CH_3OH}$=2.3, while $\gamma_{H_2O}$=2.2~\citep[][]{vanDishoeck_ea06}, which makes water starting to photodissociate at somewhat smaller radii with higher visual extinction, than methanol.

Although the composition of the bulk of ice does not directly affect the abundances of gas-phase COMs in our model, it is important to make sure that adopted parameters of our model do not lead to bulk composition that contradicts the observational data. The composition of the bulk of the ice at 1.6$\times$10$^{5}$~years is presented in Figure~\ref{fgr:ice}, bottom panel. In bulk ice, solid water is the dominant species. Solid carbon monoxide is the second most abundant bulk ice constituent in the inner and outer regions of L1544 (R$\le$0.007~pc, R$\ge$0.03~pc) with an abundance of $\sim$50\% that of solid water. In the intermediate part of the cloud (0.007~pc~$\le$~R~$\le$~0.03~pc) the abundance of CO is reduced, and the second most abundant species in the bulk ice is methanol with maximum fraction of$\sim$45\% with respect to solid water. Averaged over the whole L1544, abundances of CO and CH$_{3}$OH in the bulk of ice with respect to solid water are $\sim$35\% and $\sim$28\%, respectively. Note that bulk abundance of methanol with respect to solid water is reduced in contrast to its surface value because the methanol that buried in the bulk after its formation on surface, is partly dissociated via cosmic rays.
While the abundance of solid N$_{2}$ is not inferred from observations due to low strength of fundamental N---N stretch band of N$_{2}$ in ices~\citep[][]{Sandford_ea01}, abundances of solid CO and CH$_{3}$OH have been estimated observationally. \citet[][]{Oeberg_ea10} provided average observational fractions of CO and CH$_{3}$OH in low-mass clouds to be 29\% and 3\%, respectively, which somewhat differs from our modeling results. On the other hand, the fraction of methanol in the ice vary by $\sim$ an order of magnitude among different clouds, reaching 25\% wrt. solid water in some cases, e.g. Class~0 protostars~\citep[][]{Pontoppidan_ea04}, which are close evolutionary to the prestellar cores such as L1544. As such, we believe that the CH$_{3}$OH/CO ratio obtained in the ice by our model does not represent a major issue in this study. The abundance of ammonia in the ice is $\sim$7\% wrt. water, which is similar to observed values. Solid carbon dioxide CO$_{2}$ is one of the main constituents of the ice according to observations~\citep[e.g.,][]{Oeberg_ea10}. However, in our simulations it is missing because it is likely formed at the onset of a cold dense cloud formation from a warmer translucent cloud at T$_{dust}\sim$20~$K$~\citep[][]{Mennella_ea06, GarrodPauly11, VasyuninHerbst13}. Since in our simulations we do not follow the cloud formation with time-dependent physical conditions, we miss this phase.
Note however, that protonated carbon dioxide, HOCO$^{+}$, has recently been observed in L1544~\citep[][]{Vastel_ea16}, which implies the presence of gas-phase CO$_{2}$, too. Nevertheless in the model discussed in this study, neither gas-phase nor solid CO$_{2}$ and its protonated form do not affect the chemistry of COMs. Abundances of major ice constituents in the whole core are summarized in the Table~\ref{tbl:ice}.

\subsection{Chemistry at the COMs peak}\label{chem_COMs_peak}
\begin{figure}
  \includegraphics[height=0.8\textwidth, angle=90]{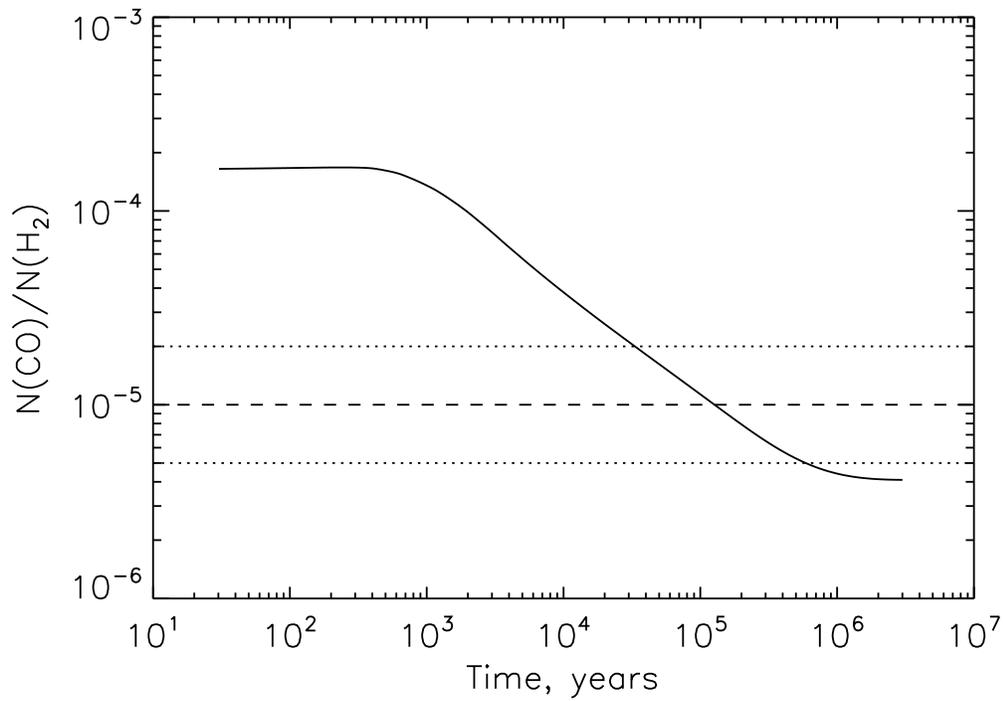}\vspace{5mm}
  \caption{CO--to--H$_{2}$ column density ratio. The value observed toward the center of L1544 is reached within 10$^{5}$ years, with our assumption of a static cloud. The observed value according to \citet[][]{Caselli_ea99} is plotted with dashed horizontal line. The factor of 2 uncertainty in the observed value is shown with the dotted horizontal lines.}
  \label{fgr:co2h2coldens}
\end{figure}

\begin{figure}
\includegraphics[width=0.80\textwidth, angle=0]{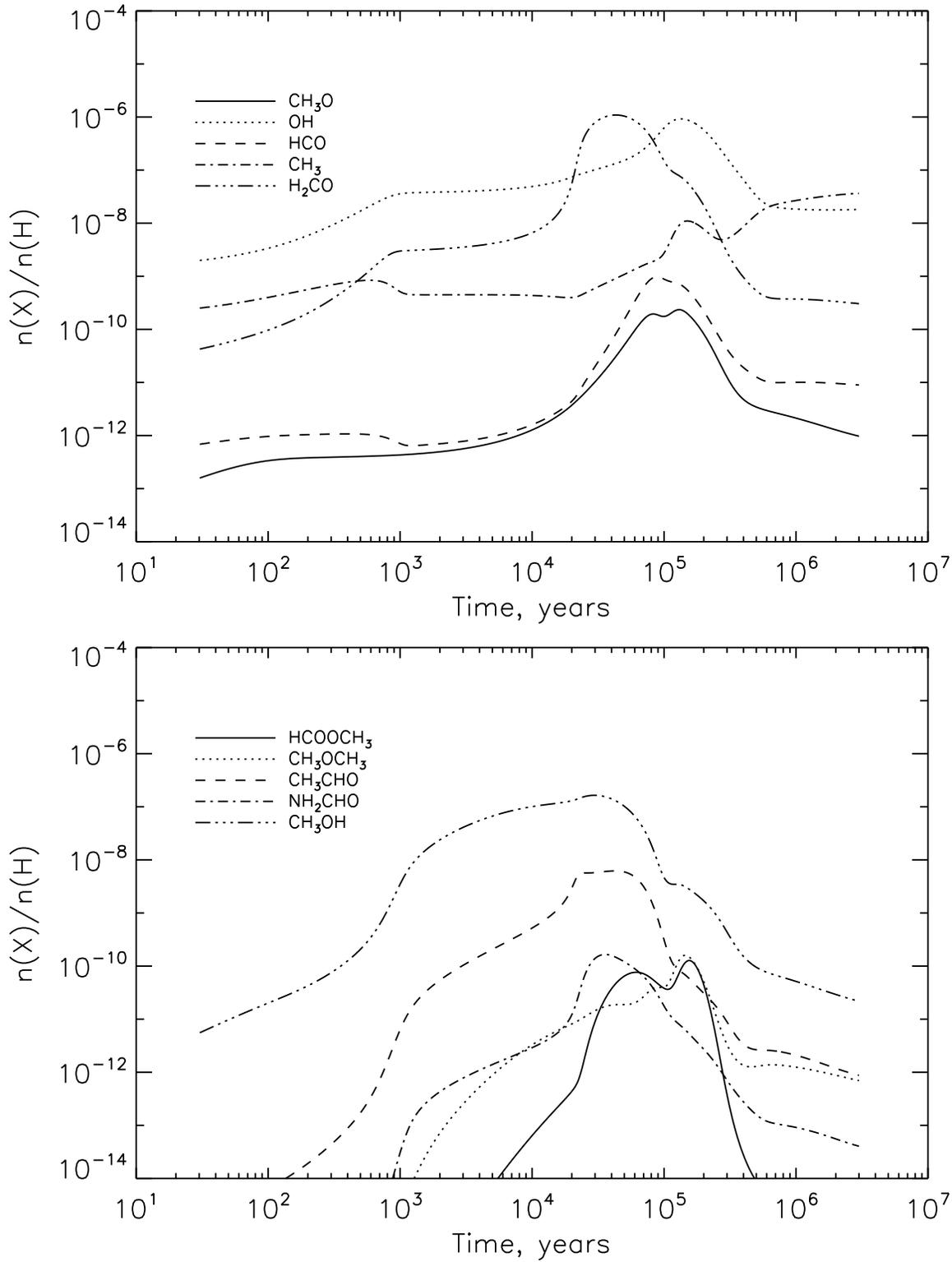}
\caption{Temporal evolution of the abundances of COMs and related species at the distance of 0.015~pc from the center of L1544. This location is chosen as representative for the chemistry at the COMs peak.}\label{fgr:coms_vs_t}
\end{figure}

\begin{figure}
\includegraphics[width=0.45\textwidth, angle=0]{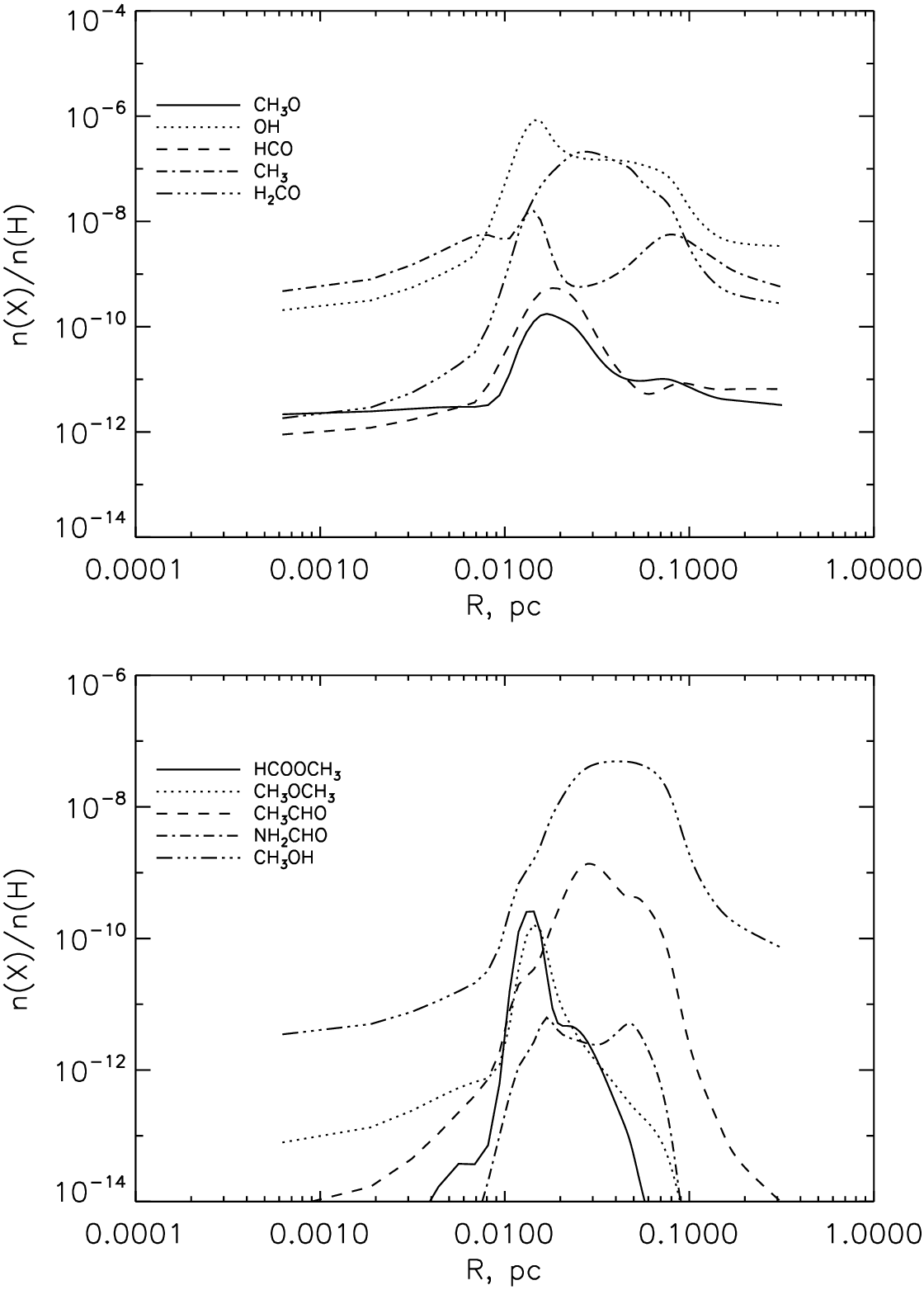}
\includegraphics[width=0.45\textwidth, angle=0]{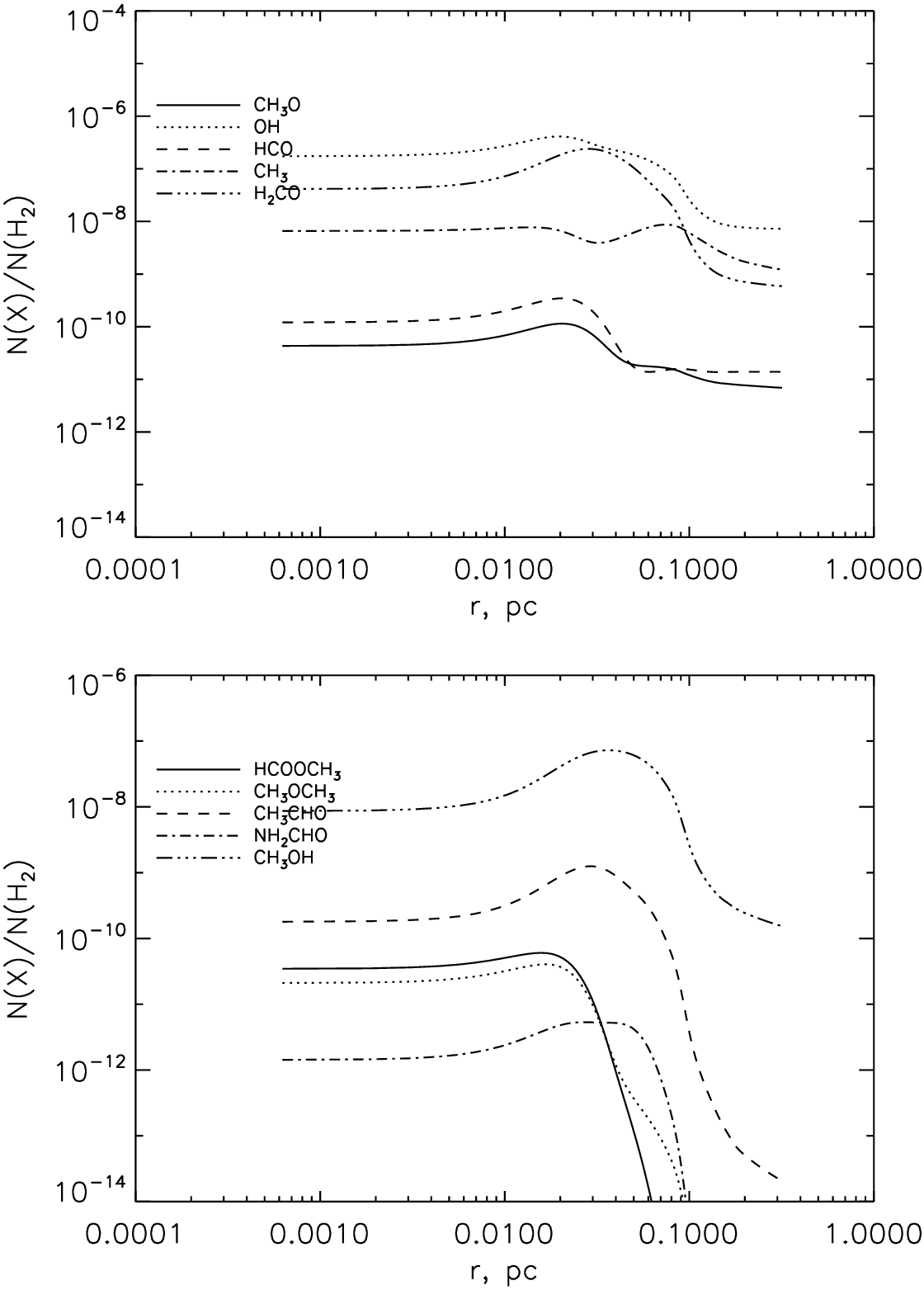}
\caption{Radial profiles of abundances of COMs and some related species at 1.6$\times$10$^{5}$~years (left column) and simulated radial profiles of abundances as obtained as column density ratios observed with 26'' telescope beam (right column). 1.6$\times$10$^{5}$~years is the time of the best agreement (minimal F(r,t), see text) between the modeled and observed abundances of species inferred from column densities. All species clearly exhibit peaks at radial distances between 0.01~pc and 0.03~pc, which matches the ring-like distribution of CH$_{3}$OH observed in L1544~\citep[][]{Bizzocchi_ea14}.}\label{fgr:multispecies_gp_coms}
\end{figure}

Modeled abundances of complex organic species in L1544 exhibit both significant temporal (Figure~\ref{fgr:coms_vs_t}, bottom panel) and radial (Figure~\ref{fgr:multispecies_gp_coms}, bottom panel) variations, as well as the abundances of chemically-related precursor species (the same figures, upper panels). The peak abundances of COMs are reached within 10$^{5}$---2$\times$10$^{5}$~years of chemical evolution, which corresponds to the time when the observed depletion factor of CO is attained in our model (Figure~\ref{fgr:co2h2coldens}). This time is higher than the time estimated by \citet[][]{Caselli_ea99} (10$^{4}$ years), using CO depletion factor and freeze-out rate. However, taking into account the uncertainty by a factor of 2 in the depletion factor claimed by \citet[][]{Caselli_ea99}, and possibly smaller depletion rate due to non-thermal evaporation of species, which was not considered by \citet[][]{Caselli_ea99}, we believe that our results do not contradict observational constraints.

Maximum abundances of most COMs including HCOOCH$_{3}$, CH$_{3}$OCH$_{3}$ and NH$_{2}$CHO are reached for distances between $\sim$0.01---0.02~pc from the center of L1544, that roughly corresponds to the position of a ring-like emission of methanol observed by \citet[][]{Bizzocchi_ea14} and of COMs deduced by \citet[][]{Vastel_ea14}. As shown by \citet[][]{JimenezSerra_ea16}, COMs are indeed enhanced in an outer shell centered at a distance of 0.02 pc (equivalent to 4000 au) with respect to the center of the core. Peak abundances of methyl formate, dimethyl ether and formamide reach approximately (3---4)$\times$ 10$^{-10}$ wrt. total hydrogen nuclei. The comparison between models and observations versus time and radial distance in L1544 is shown in Figure~\ref{fgr:agreement_maps}. In the left and middle columns, an agreement for illustrative models is shown (see below). In the right column, the agreement for the Main~model~(MM) under discussion is presented. The agreement was calculated using the following expression:
\begin{equation}\label{agreement}
F(r,t)=\sum_{i=1}^{N_{species}}\left(\frac{X_{obs}^i-X_{mod}^i(r,t)}{X_{obs}^i+X_{mod}^i(r,t)}\right)^2,
\end{equation}
where X$_{mod,obs}$ are the modeled and observed column densities of species. Modeled column densities are smoothed over the gaussian beam of 26'' size in order to simulate the observed abundances and column densities taken from \citet[][]{JimenezSerra_ea16} toward the position of the ``methanol peak''\footnote{This position corresponds to the peak of the CH$_{3}$OH emission reported by \citet[][]{Bizzocchi_ea14} which is approx. 4000~AU (0.019~pc) away from the dust emission peak (see also~\citealt[][]{JimenezSerra_ea16} for more details)}. (r,t) is the location in the phase space (radius, time). The species used to calculate the agreement are HCOOCH$_{3}$, CH$_{3}$OCH$_{3}$, NH$_{2}$CHO, CH$_{3}$CHO, CH$_{3}$O and CH$_{3}$OH. As such, an agreement map shows at which time(s) and for which radii the modeled column densities fit the observational data at the ``methanol peak'' in L1544, located at $\sim$0.02~pc or $\sim$4000~AU from the center of L1544.

In Table~\ref{tbl:abu}, the best-fit modeled column density ratios of species wrt.~H$_{2}$ -- N(X)/N(H$_{2}$), corresponding to the abundances inferred from observations for L1544 -- are shown, as well as the observed abundances obtained by \citet[][]{JimenezSerra_ea16}. Also, for comparison, abundances of COMs in L1689B and B1-b are shown, taken from \citet[][]{Bacmann_ea12} and \citet[][]{Cernicharo_ea12}. These two sources have gas and dust temperatures similar to those in L1544. Also, column densities of COMs observed toward these cores are of the same order of magnitude as in L1544 ($\sim$10$^{12}$~cm$^{-2}$) suggesting similar extinctions in the regions of COMs formation, thus making reasonable comparison with model predictions originally made for L1544. The minimal value of F(r,t) is reached at 1.6$\times$10$^{5}$~years and at the radius of 0.015~pc. This time is similar to that at which the model also reproduces the CO column density (see Figure~\ref{fgr:co2h2coldens}). This indicates that chemistry of COMs clearly belongs to a so-called early chemistry type. Also, the spatial location of the best-fit position is roughly similar to the location of the ``methanol peak'' in L1544. Since the values in Table~\ref{tbl:abu} have been calculated by averaging the predicted column densities within a beam of 26'' (this is the beam size in the observations by \citet[][]{JimenezSerra_ea16}), they are somewhat different from true local fractional abundances of species obtained with the chemical model and discussed below.

Let us now focus on the chemistry of complex organic molecules in our model near the location of best-fit abundances of COMs, namely at radial distance 0.015~pc. In general, the formation of all complex organic species considered in this study follows the scenario proposed in \citet[][]{VasyuninHerbst13_coms}. In that scenario, complex organic molecules in the cold gas are formed mainly via gas-phase reactions from precursor species that are formed on grains (totally, or at least in a significant part) and then ejected into the gas phase via efficient reactive desorption. While in \citet[][]{VasyuninHerbst13_coms}, the majority of gas-phase reactions important for the formation of COMs are ion-molecule reactions, in this study new gas phase neutral-neutral reactions play an important role too.

Dimethyl ether (CH$_{3}$OCH$_{3}$) and methyl formate (HCOOCH$_{3}$) exhibit very similar radial (see Figure~\ref{fgr:multispecies_gp_coms}) and temporal (see Figure~\ref{fgr:coms_vs_t}) behavior. Dimethyl ether is produced via the reaction
\begin{equation}\label{ch3_ch3o}
{\rm CH_{3}}+{\rm CH_{3}O}\rightarrow \rm{CH_{3}OCH_{3}}+h\nu,
\end{equation}
proposed by \citet[][]{VasyuninHerbst13_coms}. Following the comment by \citet[][]{Balucani_ea15} on the rate of this reaction, we updated the value of the rate coefficient to 3.0$\times$10$^{-10}$~cm$^{3}$s$^{-1}$ (see Table~\ref{tbl:newr}). CH$_{3}$ is formed as a product of dissociative recombination of CH$_{5}^{+}$. This formation route in Main~model~(MM) is assisted by a gas-phase reaction
\begin{equation}
{\rm CH_{3}O}+{\rm H}\rightarrow {\rm CH_{3}}+{\rm OH},
\end{equation}
also introduced in \citet[][]{VasyuninHerbst13_coms}, and leads to the increase of the abundance of CH$_{3}$ by a factor of three with respect to pure ion-molecule formation. About 55\% of the methoxy radical CH$_{3}$O in the gas phase is produced via the gas-phase reaction
\begin{equation}\label{oh_ch3oh}
{\rm OH}+{\rm CH_{3}OH}\rightarrow {\rm CH_{3}O}+{\rm H_{2}O},
\end{equation}
while another 45\% of CH$_{3}$O is ejected to the gas via reactive desorption after the surface production in the reaction
\begin{equation}\label{gh_gh2co}
{\rm s-H}+{\rm s-H_{2}CO}\rightarrow {\rm s-CH_{3}O}.
\end{equation}
\citet[][]{Shannon_ea13} found reaction (\ref{oh_ch3oh}) to exhibit strong negative dependence on temperature due to the formation of a hydrogen-bonded complex that lives sufficiently long to undergo quantum mechanical tunneling through the activation barrier of the reaction and to form products. \citet[][]{Antinolo_ea16} confirmed this result, and measured the temperature-dependent rate of the reaction in the range 22~K---64~K. Extrapolation to 10~K gives the rate of the reaction equal to 1.1$\times$10$^{-10}$~cm$^{3}$s$^{-1}$, which is three times faster than the rate adopted for this reaction in \citet[][]{VasyuninHerbst13_coms}. Thus in this work, methoxy radical reaches the fractional abundance of 2.5$\times$10$^{-10}$~wrt.~H (see Figure~\ref{fgr:multispecies_gp_coms}, upper panel), which is higher than in the model by \citet[][]{VasyuninHerbst13_coms} and about an order of magnitude higher than observed by \citet[][]{BacmannFaure16} for several prestellar cores other than L1544, and by factors of $\sim$5--10 in L1544~\citep[][]{JimenezSerra_ea16}. Note however, that the CH$_{3}$O abundance in our model is mainly controlled by the pair of reactions (\ref{ch3_ch3o}) and (\ref{oh_ch3oh}). Rates of both reactions are poorly known. Thus, in principle, by varying these reaction rates, one can modify the abundance of CH$_{3}$O widely, almost without changing the abundance of CH$_{3}$OCH$_{3}$. More theoretical and/or laboratory work needs to be carried out to put more stringent constraints on chemical models.

Methyl formate (HCOOCH$_{3}$) has two main production channels. Approximately 2/3 of the total methyl formate production belongs to the reaction
\begin{equation}\label{dme_mf_link}
{\rm O}+{\rm CH_{3}OCH_{2}}\rightarrow {\rm HCOOCH_{3}}+{\rm H}.
\end{equation}
This reaction has been studied by \citet[][]{HoyermannNacke96} and \citet[][]{Song_ea05}, and introduced into astrochemical models by \citet[][]{Balucani_ea15}. The fractional abundance of atomic oxygen in our model after 10$^{5}$~years is 6$\times$10$^{-8}$ wrt. H. The second reactant, CH$_{3}$OCH$_{2}$, is mainly produced via the reactions
\begin{equation}\label{ch3och3_oh}
{\rm CH_{3}OCH_{3}}+{\rm OH}\rightarrow {\rm CH_{3}OCH_{2}}+{\rm H_{2}O}.
\end{equation}
This reaction has been studied in \citet[][]{Shannon_ea14}, who showed that it is efficient at low temperatures due to a similar mechanism as that for reaction~(\ref{oh_ch3oh}). This is in contrast to \citet[][]{Balucani_ea15}, who pointed out the existence of a chemical link between CH$_{3}$OCH$_{3}$ and HCOOCH$_{3}$ via the reaction (\ref{dme_mf_link}), but claimed the main formation routes of CH$_{3}$OCH$_{2}$ to be reactions between dimethyl ether and elemental fluorine (F) and clorine (Cl). In our model, both chemical elements are highly depleted in the gas phase, having fractional abundances of 2.3$\times$10$^{-17}$ and 6.3$\times$10$^{-11}$ wrt.~H, respectively, at the time of 10$^{5}$~years. Therefore, the formation of CH$_{3}$OCH$_{2}$ in our model is dominated by reaction (\ref{ch3och3_oh}) instead of by reactions between dimethyl ether and F or Cl.

From this analysis, the hydroxyl radical (OH) appears to be the key chemical species for the chemistry of methyl formate and dimethyl ether in the cold gas. The fractional abundance of OH reaches the value of $\sim$7$\times$10$^{-7}$ after 10$^{5}$~years of evolution and at radial distance R=0.015~pc. This is somewhat higher than the OH abundance estimated by \citet[][]{Harju_ea00} who found it to be about 1$\times$10$^{-7}$ wrt. H. However, taking into account the intrinsic uncertainties in astrochemical models~\citep[][]{Vasyunin_ea04, Wakelam_ea06, Vasyunin_ea08, Wakelam_ea10} and observations, one can conclude that this difference is not significant.

Approximately 1/3 of total methyl formate production belongs to the ion-neutral route that includes the formation of protonated methyl formate via the reaction
\begin{equation}\label{mf_im}
{\rm CH_{3}OH_{2}^{+}}+{\rm HCOOH}\rightarrow {\rm HC(OH)OCH_{3}^{+}}+{\rm H_{2}O},
\end{equation}
with subsequent dissociative recombination into HCOOCH$_{3}$. Reaction (\ref{mf_im}) has been turned down in \citet[][]{Horn_ea04} who cited an older study by \citet[][]{Freeman_ea78}. However, in a recent study by \citet[][]{Cole_ea12}, a 5\% probability for the formation of a protonated methyl formate (with the rest 95\% probability to form an adduct ion) in reaction~(\ref{mf_im}) has been found. A protonated methyl formate is converted in our model into HCOOCH$_{3}$ in a dissociative recombination reaction with electrons with a probability of 50\%. However one should note, that the latter probability is poorly known and could be much smaller~\citep[][]{Vigren_ea10}, resulting in much lower production of methyl formate by the considered ion-molecular route.

Formic acid (HCOOH), second reactant in reaction (\ref{mf_im}), is another brick of organic chemistry in star-forming regions. It has also been detected by \citet[][]{Vastel_ea14} towards L1544, with fractional abundance of $\sim$10$^{-10}$~wrt.~H. In our model, the gas-phase abundance of formic acid towards the location of the ``methanol peak'' at the time of best agreement (1.6$\times$10$^{5}$~years) is 2.0$\times$10$^{-10}$. HCOOH is formed as a product of dissociative recombination of its protonated form HCOOH$_{2}^{+}$, which is, in turn, formed in a slow radiative association reaction HCO$^{+}$~+~H$_{2}$O~$\rightarrow$~HCOOH$_{2}^{+}$~+~$h\nu$. The surface formation route of formic acid in the reaction between OH and HCO is inefficient because in our model both reactants do not diffuse in the surface at the low temperatures of pre-stellar cores. Nevertheless, a total abundance of HCOOH of $\sim$7$\times$10$^{-9}$~wrt.~H is accumulated in the ice via gas accretion. Compilation done by \citet[][]{GarrodHerbst06} shows that the observed abundances of HCOOH in hot cores vary in the range [8.0$\times$10$^{-10}$~---~6.2$\times$10$^{-8}$]~wrt.~H. Thus, assuming that hot core/corino is the next stage in the development of a protostar that inherits the chemical composition from the previous stage, one may deduce that the total production of HCOOH in our model is within reasonable limits.

Nitrogen-bearing complex organic molecules with a peptide bond such as formamide (NH$_{2}$CHO) exhibit a plateau-like abundance profile between 0.01~pc and 0.05~pc at 10$^{5}$~years, reaching $\sim$6$\times$10$^{-12}$~wrt.~H (see Figure~\ref{fgr:multispecies_gp_coms}, bottom panel). Formamide in our model is solely produced via reaction (\ref{nh2_h2co}) despite the existence of two surface reactions, leading to NH$_{2}$CHO. Those are s-N+s-CH$_{2}$OH$\rightarrow$ s-NH$_{2}$CHO and s-NH$_{2}$+s-HCO$\rightarrow$ s-NH$_{2}$CHO. At low dust temperatures of $\sim$10~K, typical for L1544, those reactions are not efficient in our model, because with the adopted diffusion-to-desorption energy ratio E$_{b}$/E$_{D}$=0.55, the grain surface diffusion timescale even for relatively weakly bound nitrogen atoms exceeds 10$^{13}$~years.

Both reactants in reaction (\ref{nh2_h2co}), NH$_{2}$ and H$_{2}$CO, are produced in the gas phase (20\% of NH$_{2}$, 80\% of H$_{2}$CO) and on surface ice, and then ejected non-thermally to the gas. Amidogen (NH$_{2}$) on surface is produced in the hydrogenation reaction H~+~NH~$\rightarrow$~NH$_{2}$, and in the gas as a product of dissociative recombination of NH$_{4}^{+}$. Formaldehyde (H$_{2}$CO) on surface is produced in the hydrogenation reaction H+HCO$\rightarrow$H$_{2}$CO, and in the gas phase via the reaction O~+~CH$_{3}~\rightarrow$~H${_2}$CO~+~H. The possible reason why NH$_{2}$CHO is somewhat overproduced in our model with respect to the observed upper limit (see Table~\ref{tbl:abu}) is due to too high efficiencies of formation of the reactants of the reaction (\ref{nh2_h2co}). See Section~\ref{section:discussion} for details.

Acetaldehyde (CH$_{3}$CHO) is mainly produced in reaction~(\ref{ch_ch3oh_1}). Due to the high rate of this reaction, both temporal and spatial profiles of acetaldehyde are well correlated with those of methanol, and CH$_{3}$CHO reaches peak abundances about two orders of magnitude higher than other COMs considered in this study, although at very early time (see Figure~\ref{fgr:coms_vs_t}). By the time of the best agreement with observations, 1.6$\times$10$^{5}$~years, the abundance of CH$_{3}$CHO is within an order of magnitude to those of HCOOCH$_{3}$ and CH$_{3}$OCH$_{3}$ (see Figure~\ref{fgr:coms_vs_t} and Figure~\ref{fgr:multispecies_gp_coms}), as well as to its observed value (see Table~\ref{tbl:abu}).

This mechanism of formation of acetaldehyde is different from that described in \citet[][]{VasyuninHerbst13_coms}, where acetaldehyde was mainly produced in the gas phase via the neutral-neutral reaction O~+~C$_{2}$H$_{5}~\rightarrow$~CH$_{3}$CHO~+~H. The reason for this is the updated treatment of reactive desorption in the current study. In \citet[][]{VasyuninHerbst13_coms}, C$_{2}$H$_{5}$ is delivered to the gas from the surface via reactive desorption. In \citet[][]{VasyuninHerbst13_coms}, all species formed on surface had single reactive desorption efficiency of 10\%, while in the current study, we use expression (\ref{rdeff}). For the surface reaction s-H+s-C$_{2}$H$_{4}\rightarrow$C$_{2}$H$_{5}$ which is the main source of C$_{2}$H$_{5}$ in \citet[][]{VasyuninHerbst13_coms}, expression (\ref{rdeff}) gives the efficiency of reactive desorption 0.0005\%, even for the part of surface consisting of heavy species, i.e., effectively zero. As such, the chemistry of acetaldehyde becomes different in this study in comparison to \citet[][]{VasyuninHerbst13_coms}, and the peak abundance of CH$_{3}$CHO in the absence of reaction (\ref{ch_ch3oh_1}) becomes two orders of magnitude lower than in \citet[][]{VasyuninHerbst13_coms}.

Finally, let us consider the chemistry of methanol in the gas phase in our model of L1544. There are no significant contributions from gas-phase reactions to the formation of methanol. The full amount of methanol is ejected to the gas phase from the surfaces of icy mantles of interstellar grains, where CH$_{3}$OH is formed as a product of hydrogenation of CO molecule (see Section~\ref{section:methanol_formation}). The main ejection mechanism is non-thermal reactive desorption from the part of the surface covered with ``heavy species'' (see Figure~\ref{fgr:heavy-light-ice}), with an efficiency of 0.64\% that matches previous more conservative estimates of reactive desorption efficiency~\citep[][]{Garrod_ea07}. Other non-thermal desorption mechanisms (photodesorption and cosmic ray--induced desorption) do not contribute significantly to the desorption rate of methanol. The main destruction route of methanol in the gas phase is reaction (\ref{oh_ch3oh}), as well as ion-molecule reactions with major ions such as H$_{3}^{+}$ and H$_{3}$O$^{+}$. It is important to note that the peak local volume abundances of COMs in our model does not coincide with the peak local volume abundance of methanol (see Figure~\ref{fgr:multispecies_gp_coms}, left lower panel), although due to averaging effects in single-dish observations, the peak abundances of COMs and methanol inferred from column densities spatially coincide near the observed ``methanol peak'' (see Figure~\ref{fgr:multispecies_gp_coms}, lower right panel). As such, a very high abundance of methanol in the gas phase is not needed to reproduce observed abundances of COMs. Moreover, we attribute the highest abundance of methanol reached in its radial profile (7$\times$10$^{-8}$~wrt.~H at R=0.04~pc) to the fact that dynamics is ignored. In a dynamical model, which starts with lower densities, the CO freeze out will be less efficient and, consequently, less CH$_{3}$OH is expected to be produced.

\subsection{The role of reactive desorption in the chemistry of cold COMs}\label{secton:rd_role}
\begin{figure}
  \includegraphics[width=0.45\textwidth, angle=0]{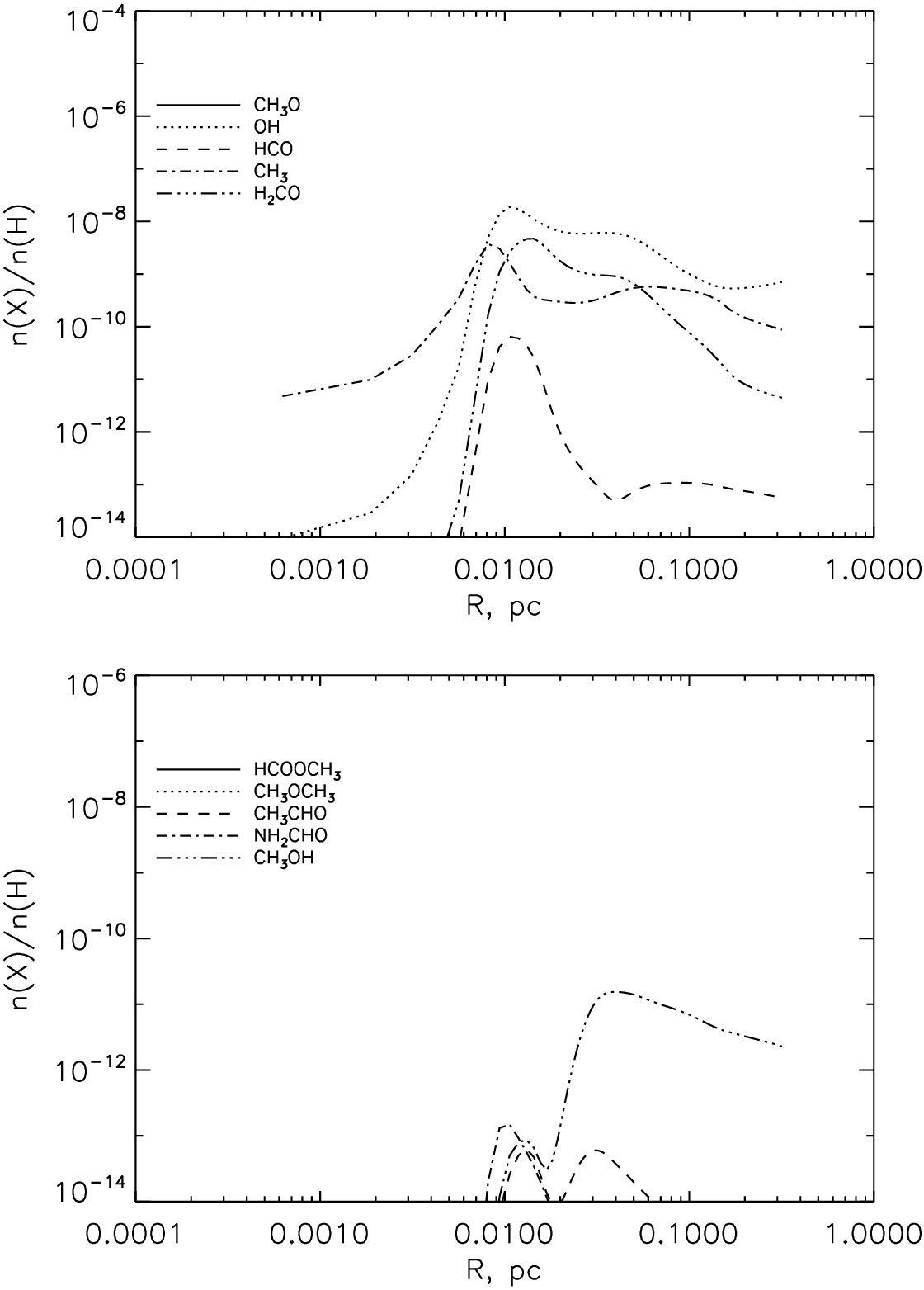}
  \includegraphics[width=0.45\textwidth, angle=0]{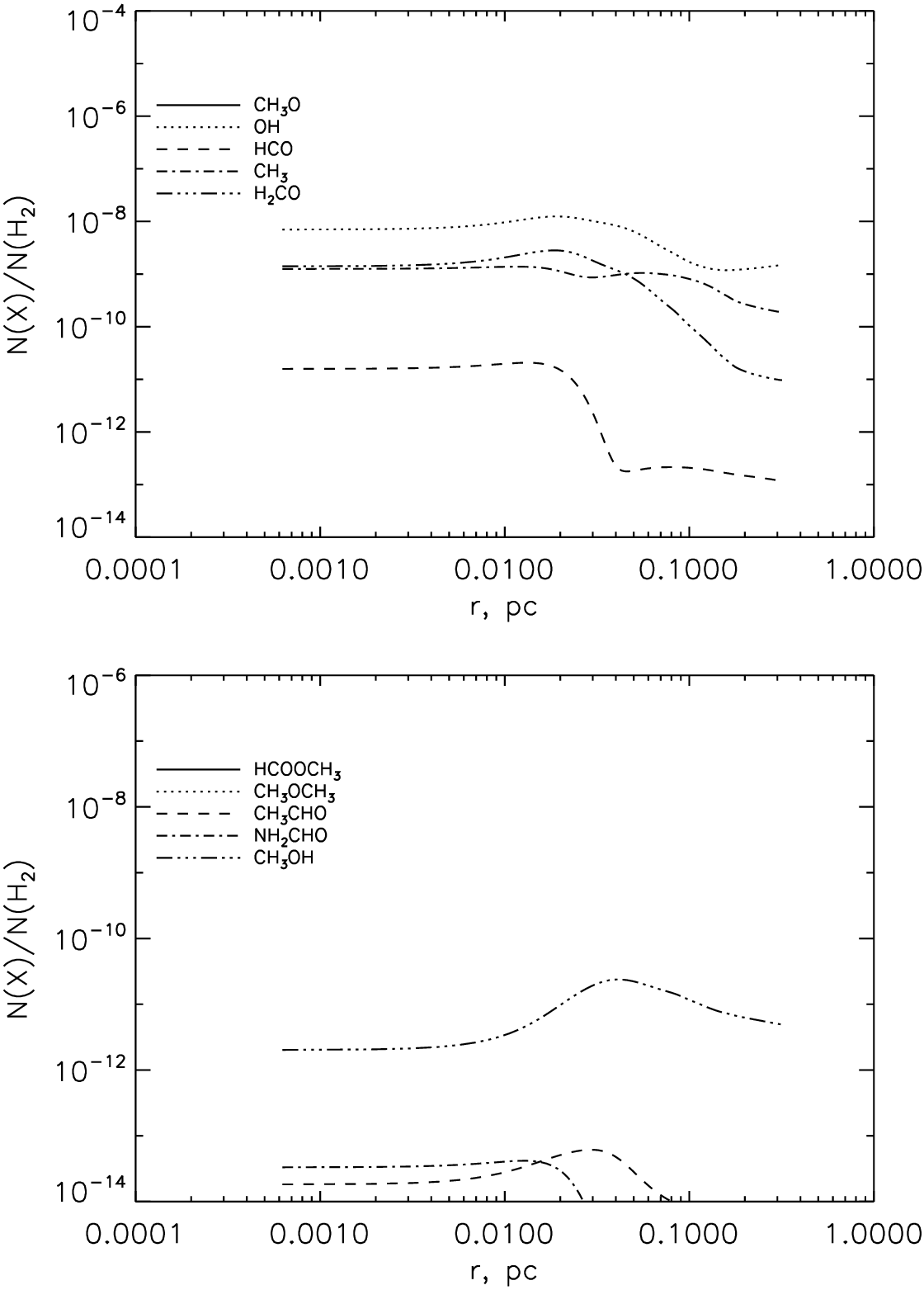}
  \caption{Same as in Figure~\ref{fgr:multispecies_gp_coms}, but at 8.0$\times$10$^{4}$~years calculated with reactive desorption completely disabled. 8.0$\times$10$^{4}$~years is the time of the best agreement (minimal F(r,t), see text) between the modeled and observed abundances of species inferred from column densities. New gas-phase chemical reactions are not sufficient to reproduce abundances of COMs without supply of precursor species from grain surface via reactive desorption.}
  \label{fgr:alternate_rd_no}
\end{figure}

\begin{figure}
  \includegraphics[width=0.45\textwidth, angle=0]{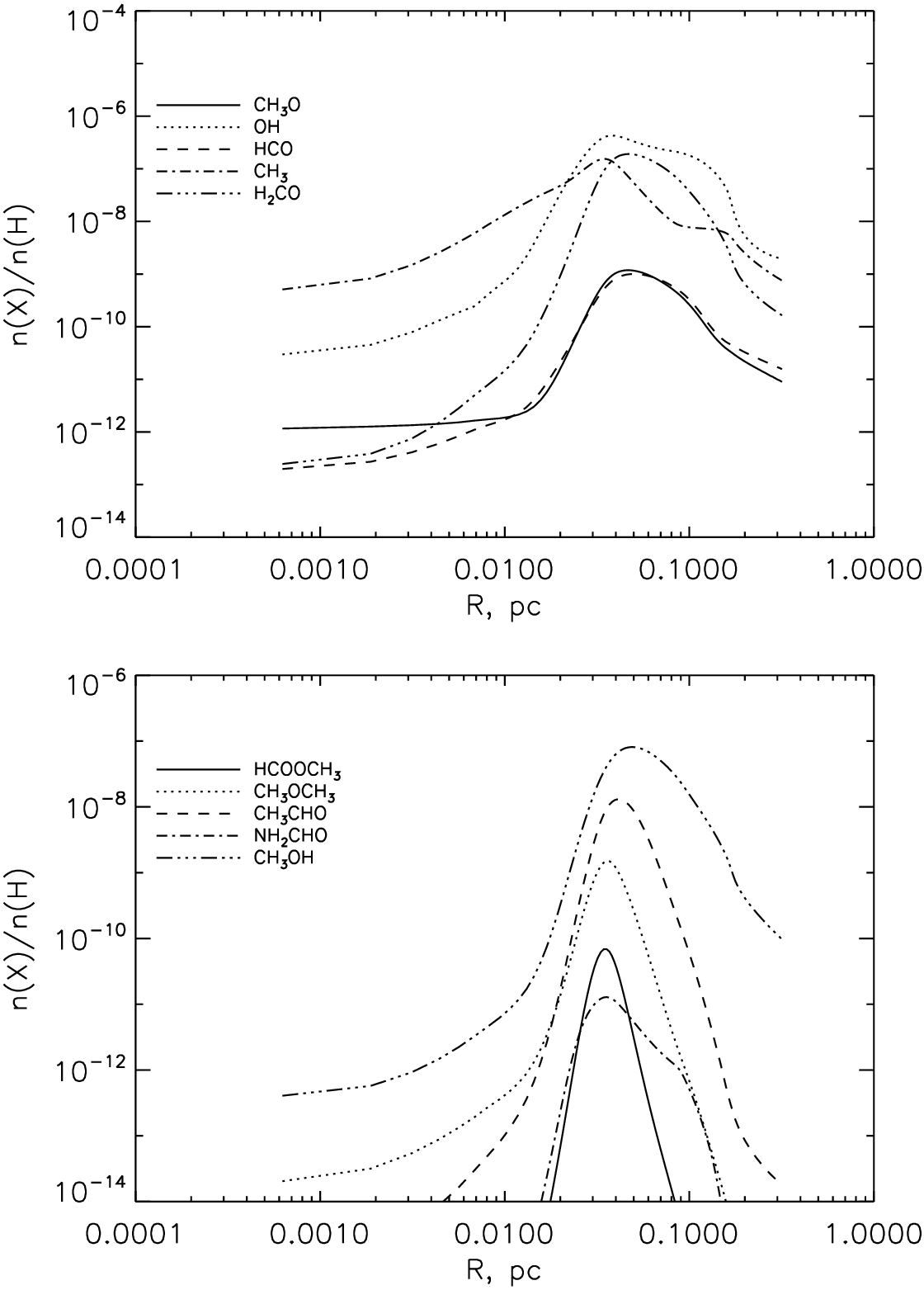}
  \includegraphics[width=0.45\textwidth, angle=0]{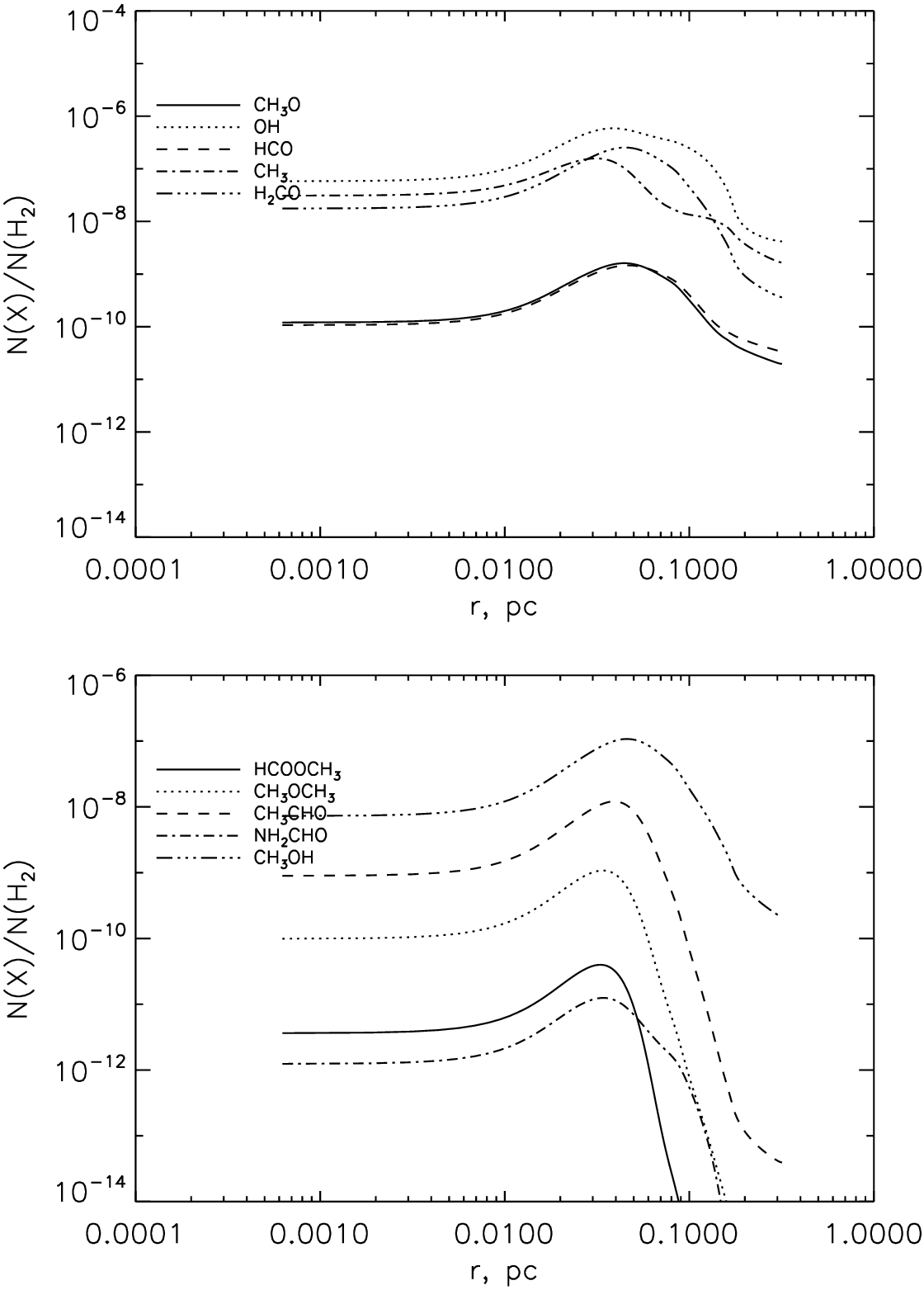}
  \caption{Same as in Figure~\ref{fgr:multispecies_gp_coms}, but at 3.0$\times$10$^{6}$~years calculated with single 10\% efficiency of reactive desorption. 3.0$\times$10$^{6}$~years is the time of the best agreement (minimal F(r,t), see text) between the modeled and observed abundances of species inferred from column densities.}
  \label{fgr:alternate_rd_low}
\end{figure}

Although in our model complex organic molecules in the cold environments typical of prestellar cores are formed via gas-phase chemistry, the role of surface chemistry in the formation of precursor species, and of reactive desorption for their delivery to the cold gas, is pivotal. To illustrate this, we ran two models with different efficiencies of reactive desorption. In the first illustrative model (IM0), reactive desorption is switched off (efficiency 0\%). In the second illustrative model (IM10), the reactive desorption efficiency for all species is 10\%, same as in \citet[][]{VasyuninHerbst13_coms}. Best-fit radial profiles of abundances of COMs and their precursors for the illustrative models are shown in Figure~\ref{fgr:alternate_rd_no} for the model with no reactive desorption and in Figure~\ref{fgr:alternate_rd_low} for the model with 10\%-efficiency reactive desorption. In addition, in Table~\ref{tbl:surface_contribution}, we list the main precursors of COMs in the gas phase, their reactive desorption efficiencies and contribution from surface chemistry to the total rate of formation rates of those species in the gas phase in our main model with advanced treatment of reactive desorption.

\begin{figure}
  \includegraphics[width=0.33\textwidth, angle=0]{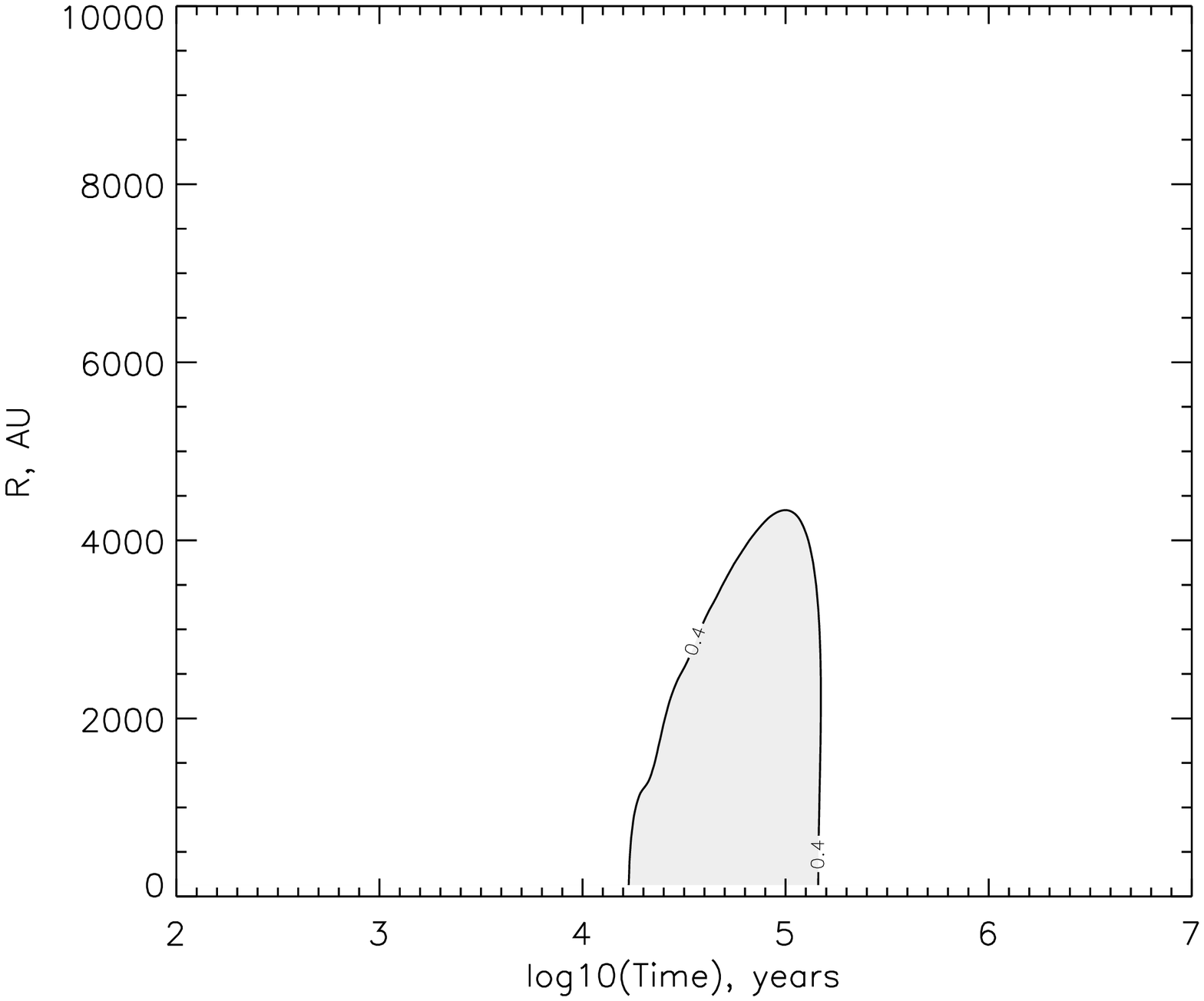}\includegraphics[width=0.33\textwidth, angle=0]{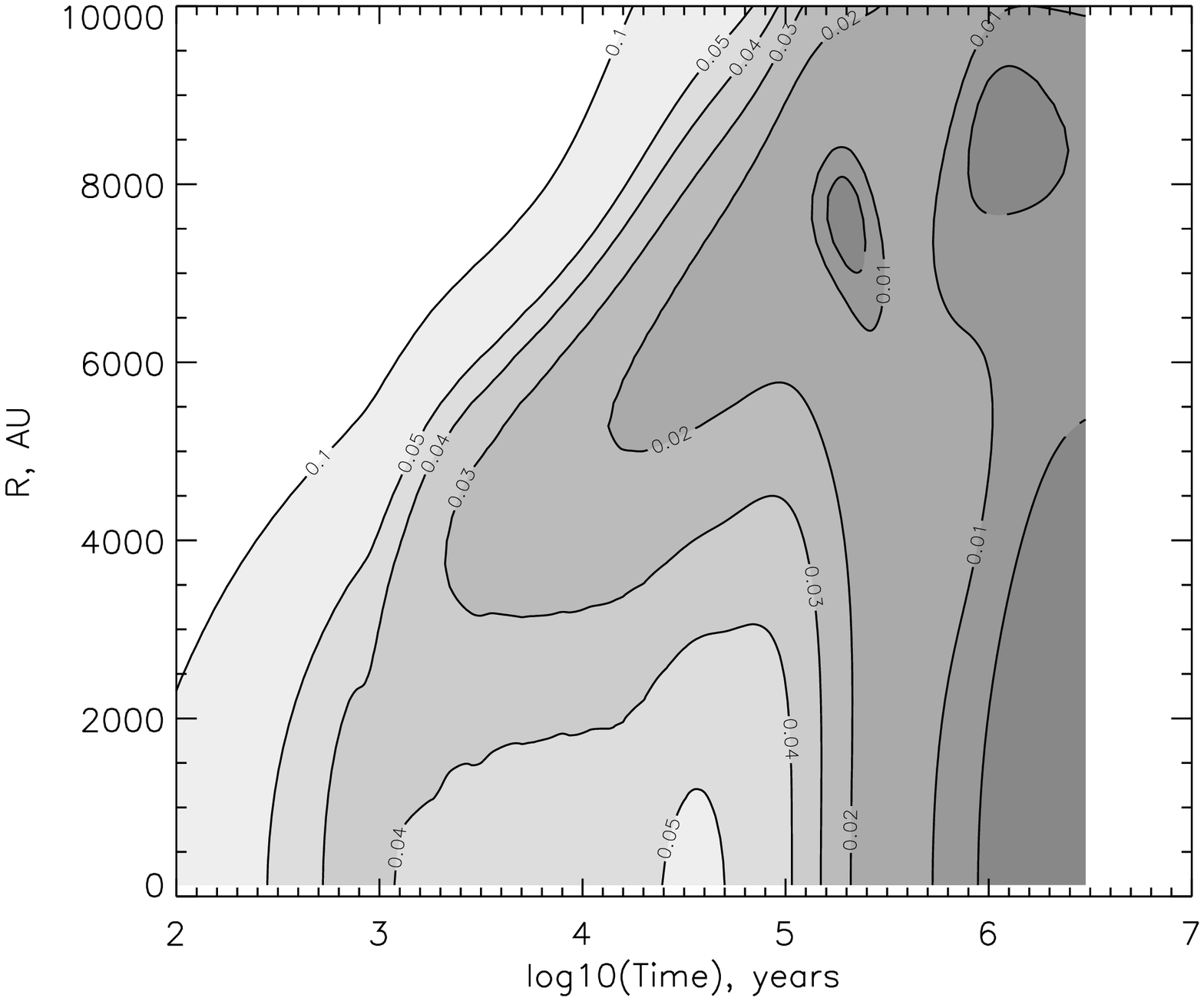}\includegraphics[width=0.33\textwidth, angle=0]{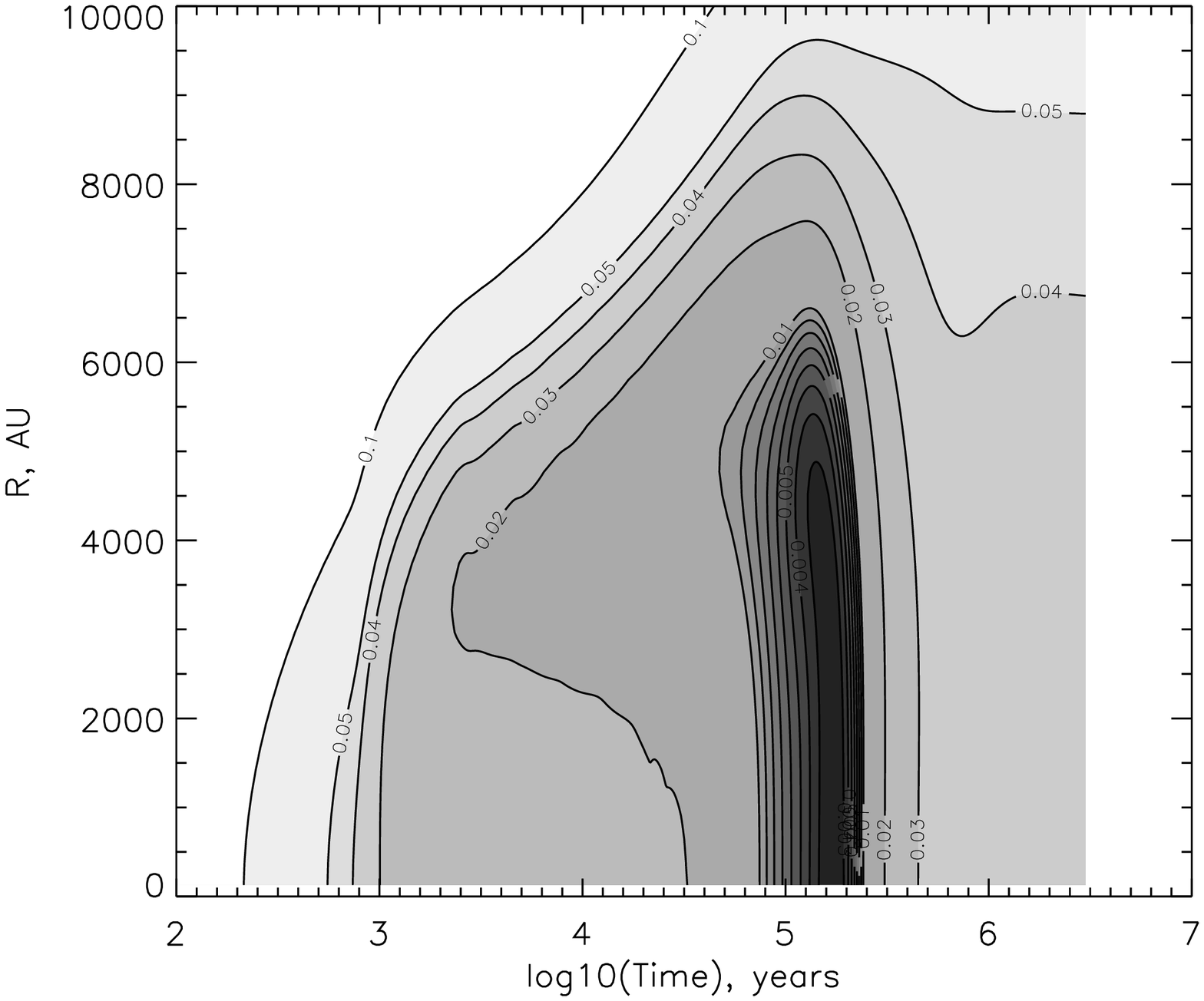}
  \caption{Agreement maps for column densities of COMs for the model with no reactive desorption (left column), model with single 10\% efficiency of reactive desorption (middle column), and model with advanced treatment of reactive desorption (right column). Modeled column densities used for comparison are smoothed over the 26'' gaussian beam. Darker filling indicates the better agreement between model and observations. Only the model with advanced treatment of reactive desorption exhibit agreement with observations for both abundances and column densities of COMs at the time corresponding to the observed amount of CO freeze-out, and radial location similar to the observed location of the ``methanol peak''. }
  \label{fgr:agreement_maps}
\end{figure}

Let us first consider the illustrative model with reactive desorption switched off (IM0). As can be seen on the ``agreement map'' in Figure~\ref{fgr:agreement_maps} (left column), in the entire space (time, radius) there is no area where model matches observational data satisfactorily. As an example, the radial profiles of fractional abundances are plotted in Figure~\ref{fgr:alternate_rd_no} at 8$\times$10$^{4}$~years of evolution, that corresponds to the minimum value of F(r,t) in Eq.~(\ref{agreement}). In model IM0, fractional abundances of all COMs considered in this study do not exceed 10$^{-12}$~wrt.~H. Note that there are two other types of non-thermal desorption included in all models considered in this study --- photodesorption and cosmic ray--induced desorption. However, it appears that the efficiency of these two other types of non-thermal desorption processes are clearly not high-enough to provide sufficient amounts of precursors of COMs to the cold gas.

The resulting radial profiles of abundances obtained with the second illustrative model (IM10), in which the reactive desorption efficiency for all species is set to a single value of 10\%, are shown in Figure~\ref{fgr:alternate_rd_low} for the times of the best agreement with observations, which is 3$\times$10$^{6}$~years of cloud evolution. At that time, the modeled CO column density is too low in comparison to the observed value (see Figure~\ref{fgr:co2h2coldens}). As can be seen in Figure~\ref{fgr:agreement_maps} (middle column), the illustrative model with 10\% efficiency of reactive desorption also does not agree with observations in the entire parameter space considered (time, radius). One can see that the peak abundances of COMs in the illustrative model with reactive desorption efficiency 10\% are significantly higher than in our main model, with the exception of methyl formate. The peak abundances of COMs are located near R=0.04~pc, which is somewhat inconsistent with the observed location of the ``methanol peak'' (see \citet[][]{Bizzocchi_ea14} and \citet[][]{JimenezSerra_ea16}). The abundance of acetaldehyde (CH$_{3}$CHO) is equal to 1.0$\times$10$^{-8}$~wrt.~H, which is higher than the abundance observed towards the hot core in Sgr~B2~\citep[][]{Belloche_ea13, Occhiogrosso_ea14}. Abundances of other species are also higher than in our main model by about an order of magnitude, with the exception of methyl formate, whose abundance drops with time faster than that of other species. In summary, for the majority of species, the IM10 model gives the values of column densities of COMs higher than those obtained with our Main~model~(MM). Averaged column densities are also higher than observed by 1--3 orders of magnitude even at the late evolutionary time of 3$\times$10$^{6}$~years. No location is found in the parameter space (time, radius) with satisfactory agreement with observations of COMs in L1544.

\section{Discussion}\label{section:discussion}
Although a remarkable progress has been made during the last years, our knowledge on the formation and evolution of organic matter in star-forming regions is still far from being comprehensive. In particular, models of formation of saturated complex organic molecules that have been found in cold clouds representing the earliest stages of star formation \citep[][]{Oeberg_ea10, Bacmann_ea12, Cernicharo_ea12, Vastel_ea14}, are under active development. As such, certain controversy exists between the models and adopted physical parameters. The Main~model~(MM) presented in this work, is a development of the model proposed in \citet[][]{VasyuninHerbst13_coms}, but it differs in treatment of reactive desorption, in the adopted set of gas-phase reactions and in the utilized parameters of surface chemistry. It is important to assess the importance of introduced changes, to compare the proposed model with others available, and to discuss the issues that exist in the model.

In contrast to \citet[][]{VasyuninHerbst13_coms} where grain surface chemistry were treated in a simplistic way without taking into account the ice thickness, in the current study, we use a multilayer approach to the chemistry on interstellar grains. Under the conditions of prestellar cores, icy mantles can reach thickness up to several hundreds of monolayers, while accumulating over large periods of time. Therefore, it is reasonable to distinguish between surface layers of ice, where the fast diffusive chemistry occurs as well as thermal and non-thermal desorption, and the inner parts of icy mantles, that may have different composition, and conditions for chemistry to occur. Following \citet[][]{Fayolle_ea11} and \citet[][]{VasyuninHerbst13}, four upper layers of the ice in our model comprise the ``surface ice''. As, according to our model, the total ice thickness in the central dark region of L1544 exceeds 200 monolayers, only a few percent of atoms and molecules residing on grains are available to participate in reactions on surface ice and be ejected to the gas via reactive desorption. This is in contrast to \citet[][]{VasyuninHerbst13_coms}, where all species on grains were available for these processes. Effectively, it means that rates of reactions causing reactive desorption in our updated model are generally smaller due to smaller number of available reactants. As such, while in \citet[][]{VasyuninHerbst13_coms}, the best-fit model has only thermal hopping as a source of diffusion of species on surface, in this study, we had to enable quantum tunneling for H and H$_{2}$ in order to ensure sufficient rates of reactions in surface layers of ice, and thus rates of reactive desorption. Therefore, our updated model with multilayer approach to ice chemistry, and advanced treatment of reactive desorption based on \citet[][]{Minissale_ea16}, requires efficient quantum tunneling for H and H$_{2}$ to reproduce the COM observations by \citet[][]{JimenezSerra_ea16} toward L1544. With tunneling for diffusion enabled, at the low temperature of $\sim$10~K, the exact value of another poorly known parameter that controls surface chemistry, diffusion-to-desorption energy ratio, becomes unimportant.

Another key assumption in this study, is the high efficiency of reactive desorption from the fraction of surface ice covered by species with molecular masses higher than those of water. The two most abundant heavy species in this study are carbon monoxide (CO) and methanol (CH$_{3}$OH). They have similar molecular masses about twice higher than the mass of water molecule. Near the location of the ``methanol peak'' in L1544, these two molecules comprise about 50\% of the surface ice composition in Main~model~(MM). Here, the importance of the multilayer approach to ice chemistry is revealed: while the bulk of ice is dominated by water, the surface layers of ice in a prestellar core are apolar, and covered with CO and methanol, which is in line with ice observations~\citep[][]{Oeberg_ea11}. While the high efficiency of reactive desorption in certain chemical reactions on bare olivine surface was confirmed experimentally \citep[][]{Dulieu_ea13}, as well as that water ice surface severely reduces the efficiency of RD \citep[][]{Congiu_ea09, Minissale_ea16}, to the best of our knowledge, there are no experimental studies of the efficiency of RD on CO and methanol ices. As such, while our assumption of higher reactive desorption efficiency from surface covered with heavy molecules (in comparison to water molecules) looks reasonable to us, it definitely requires experimental confirmation. Interestingly, in a recent study \citet[][]{Wakelam_ea17} argues that binding energies of species to water surface are generally higher than it was believed. This may further inhibit the efficiency of reactive desorption of species from water ice. However, conclusions made by \citet[][]{Wakelam_ea17} clearly do not affect results presented in the current study, since we only assume efficient reactive desorption from non-water fraction of ice surface.

It is also worth noting that the efficiency of reactive desorption in our study is determined by expression (\ref{rdeff}), which is semi-empirical in nature, and is derived based on a limited set of experiments. Expression (\ref{rdeff}) gives a wide range of RD efficiencies depending on a particular reaction and on reactants. In this study, expression (\ref{rdeff}) is also applied to systems not studied experimentally. Also, one should bear in mind that in case of certain species, expression (\ref{rdeff}) includes poorly known parameters. For example, to the best of our knowledge, the binding energy of NH$_{2}$ has never been measured. Thus, the adopted value of desorption energy for NH$_{2}$ of 3960~K may be too low, leading to the too high efficiency of reactive desorption according to (\ref{rdeff}), which is equal to 25\% (see Table~\ref{tbl:surface_contribution}). This may lead to marginally overestimated abundance of NH$_{2}$CHO in our model over the upper limit set by the observations of~\citet[][]{JimenezSerra_ea16}. On the other hand, the fractional abundance of NH$_{2}$ in our Main~model~(MM) at 10$^{5}$~years is 1.5$\cdot$10$^{-10}$ wrt.~H, which is consistent to the model of \citet[][]{LeGal_ea14}. Moreover, the NH$_{2}$:NH$_{3}$ ratio in our Main~model~(MM) is close to 1:20, which is also consistent with \citet[][]{LeGal_ea14}. Therefore, the overproduction of gas-phase H$_{2}$CO seems to be the most probable reason for the slight overproduction of NH$_{2}$CHO.

While non-thermal reactive desorption in our model is a key process that delivers precursors of COMs to the cold gas, the COMs themselves are formed via gas-phase chemical reactions. During the last decade, the long-standing paradigm of astrochemistry stating that the most important reactions in the cold ISM are ion-molecule reactions, has been somewhat changed. Since \citet[][]{Smith_ea04}, it is recognized that fast neutral-neutral reactions can affect the abundances of exotic carbon-bearing species in cold clouds. \citet[][]{VasyuninHerbst13_coms} proposed that reactions of radiative association may be important for the formation of some terrestrial-type organic molecules such as dimethyl ether. However, the majority of gas-phase reactions responsible for the formation of COMs in \citet[][]{VasyuninHerbst13_coms} are ion-molecule reactions leading to the formation of protonated COMs that must recombine with electrons in order to form neutral species. This scheme is somewhat problematic, because the outcome of neutral COMs in the recombination process is not clear.

In this study, gas-phase chemistry of COMs has more solid basis thanks to theoretical and experimental studies during the last years. The chemical link between CH$_{3}$OCH$_{3}$ and HCOOCH$_{3}$ via the intermediate species CH$_{3}$OCH$_{2}$ proposed in \citet[][]{Balucani_ea15}, solves the problem of underproduction of methyl formate via ion-molecule route discussed in \citet[][]{VasyuninHerbst13_coms}. However it is worth noting that in our study, CH$_{3}$OCH$_{2}$ is formed via reaction (\ref{oh_ch3och3}), which is proposed in \citet[][]{Shannon_ea14}, but missing in \citet[][]{Balucani_ea15}. Reactions with chlorine and fluorine in our model cannot produce sufficient amount of CH$_{3}$OCH$_{2}$ due to strong depletion of Cl and F in the cold gas. As such, it is not entirely clear to us how the model by \citet[][]{Balucani_ea15} works in the presented form.

As shown in Section~\ref{chem_COMs_peak}, the mechanism of formation of CH$_{3}$CHO discussed in \citet[][]{VasyuninHerbst13_coms} is not working in the updated model due to the new description of reactive desorption. As such, new chemical pathways to acetaldehyde in the cold gas should be considered. In this work, we included reaction (\ref{ch_ch3oh_1}) as one of the possible efficient routes to form acetaldehyde in the cold gas. It is interesting to note that methylidyne radical (CH) has certain similarity of chemical properties with the hydroxyl radical (OH). The enhanced rates of gas-phase reactions with hydroxyl radical (OH) found by \citet[][]{Shannon_ea13, Shannon_ea14} are shown to impact chemistry of complex organic molecules in cold gas. Thus, it may be reasonable to expect that reactions with methylidyne radical could have broader impact on the chemistry of COMs than currently assessed, and worth further detailed studies.
Another reaction to form CH$_{3}$CHO proposed in this work is the radiative association reaction (\ref{ch3_hco}). We propose this reaction by analogy with reaction (\ref{ch3_ch3o}), which is currently considered as a major route of the formation of dimethyl ether~\citep[][]{Cernicharo_ea12, VasyuninHerbst13_coms, Balucani_ea15}. However, one should point out that to the best of our knowledge, reaction (\ref{ch3_hco}) has not been studied under the conditions relevant to cold molecular clouds.

It is worth noting a surprisingly good agreement between the results produced by our complex and uncertain multiparameter chemical model, and the observational data presented in \citet[][]{JimenezSerra_ea16}. Also, at the time of best agreement with observations ($\sim$10$^{5}$~years), our model reproduces reasonably well the abundance ratios for COMs-related species (with the exception of formaldehyde) observed by \citet[][]{BacmannFaure16} in other prestellar cores. Namely, they found the ratios for HCO:H$_{2}$CO:CH$_{3}$O:CH$_{3}$OH $\sim$ 10:100:1:100. As can be seen in Table~\ref{tbl:abu}, our modeled ratios are 5:480:1:270. Since the sample of prestellar cores by \citet[][]{BacmannFaure16} does not include L1544, it is not clear if the reason for the disagreement belongs to deficiency of our model, or to the different physical structure.

Possible problems in our chemical model include overproduction of gas-phase formaldehyde and methanol. As noted above, this may be caused by possibly overestimated efficiency of reactive desorption of H$_{2}$CO and CH$_{3}$OH with the expression~(\ref{rdeff}) that is semiempirical in nature and includes poorly known values. However, as can be seen in Table~\ref{tbl:surface_contribution}, reactive desorption efficiencies for these species are not extremely high (0.64\% for CH$_{3}$OH and 5.4\% for H$_{2}$CO), which is consistent with conservative estimates of RD efficiencies~\citep[e.g.,][]{Garrod_ea07}. Another possible reason for the too high abundances of methanol and formaldehyde in the gas phase in our model is their overproduction on grains that is translated into gas via reactive desorption, which also leads to the somewhat increased fractions of CH$_{3}$OH and H$_{2}$CO in the ice bulk (see lower panel of Figure~\ref{fgr:ice}, Table~\ref{tbl:ice}) as compared to observed values~\citep[][]{Oeberg_ea11}. One possible remedy for this is to take into account an evolutionary model for L1544, where dust temperature changes during collapse from typical values for diffuse clouds (of $\sim$20~K) to $\le$10~K~\citep[see][]{VasyuninHerbst13}. This will ensure formation of another major ice constituent, CO$_{2}$, which is currently missing in our static model of L1544. Formation of CO$_{2}$ will lock 10--20\% of the total carbon budget, thus reducing, among others, the abundances of H$_{2}$CO and CH$_{3}$OH. The formation efficiency of H$_{2}$CO and CH$_{3}$OH may also be reduced by considering the new scenario that includes backward hydrogen abstraction reactions~\citep[][]{Minissale_ea16c}.

Although in this study we use a time-dependent chemical model, the adopted physical model of L1544 is static. Since time scales of chemical and dynamical evolution of prestellar cores are comparable, a more realistic approach should include a dynamical model of the physical structure of L1544. In fact, the inclusion of dynamical evolution could imply a longer time scale spent by the core at lower volume densities, thus with longer freeze out time scales (as the freeze out time scale is inversely proportional to the volume density). The inclusion of dynamics will be the subject of our next paper. The possible longer freeze-out times in a dynamical model may affect chemistry in the center of the core, where with the current model, an extreme freeze-out of virtually all species is observed, including nitrogen--bearing (see Figure~\ref{fgr:simple_species_vs_r}), which is somewhat inconsistent with observations. However, another possible explanation for the ``extra'' depletion of N-bearing molecules could be that the multi-layer treatment overestimates the amount of N$_{2}$ freeze-out \citep[see also][]{Sipila_ea16}, so that further laboratory studies will be needed to put more stringent constraints on the mobility of species in the bulk and surfaces of icy mantles.

\section{Summary}\label{section:summary}
In this study, we performed chemical modeling of the formation and evolution of complex organic molecules (COMs) in the prestellar core L1544. We found that saturated COMs can be formed efficiently in L1544 up to the fractional abundances of (3---4)$\times$10$^{-10}$ wrt. total number of hydrogen nuclei via the scenario proposed in \citet[][]{VasyuninHerbst13_coms}, and further developed in this study. The development includes more detailed treatment of the reactive desorption based on experiments by \citet[][]{Dulieu_ea13} and \citet[][]{Minissale_ea16}, and an extended set of gas-phase reactions, important for the formation of complex organic molecules in the cold gas. Chemistry on interstellar grains is treated via a multilayer approach, which allows to discriminate between surface and bulk ice. Abundances of many COMs such as CH$_{3}$OCH$_{3}$, HCOOCH$_{3}$ and NH$_{2}$CHO peak at similar radial distance of ~2000---4000 AU from the core center, which is in line with recent observations of L1544 performed by \citet[][]{JimenezSerra_ea16}, \citet[][]{Bizzocchi_ea14} and \citet[][]{Vastel_ea14}. Gas-phase abundances of COMs depend on the efficiency of reactive desorption, which in turn depends on the composition of the outer monolayers of icy mantles. In prestellar cores, outer monolayers of icy mantles likely include large fraction of CO and products of CO hydrogenation with atomic weights higher than that of water, which may allow the increase of the efficiency of reactive desorption in comparison to that on water ice. As such, combination of non-thermal desorption and extended gas-phase chemistry based on recent experimental and laboratory works, provides reasonable explanation for the observed abundances of complex organic molecules in L1544, and, probably, in other similar objects.

We believe that a three-phase approach (gas--ice surface--ice bulk) to gas-grain chemistry in star-forming regions should be preferred in modeling over the two-phase approach (gas--ice) even considering all the uncertainties that currently exist in our understanding of the processes on the ice surface and within the ice. The three-phase approach reflects the fundamental fact that dynamics of chemical processes on a solid surface and within the solid body are different, and allows us to explore the astrochemical importance of this fact. While currently three-phase models include more poorly known parameters than two-phase ones, this will change with time with the advent of new experimental and theoretical studies.

The chemical model employed in this study, has revealed the key role of the hydroxyl radical (OH) in the chemistry of several important complex organic molecules -- methoxy radical, dimethyl ether and methyl formate. To date, the potential importance of this species for the chemistry of COMs is not fully recognized in the literature. To the best of our knowledge, the only systematic study of OH in prestellar cores has been performed by \citet[][]{Harju_ea00}. Given the importance of hydroxyl radical for the chemistry of COMs, as well as the possible high efficiency of the entire class of neutral-neutral reactions with OH at low temperatures~\citep[][]{Shannon_ea13, Shannon_ea14}, we believe that OH radical is an important target for future observational studies.

Finally, it is important to point out that in this study, we used several important parameters that require more accurate experimental or theoretical measurements. The key assumption of this study -- the efficient reactive desorption from surface made of CO and its hydrogenation products -- is to be checked in the lab. The binding energy of NH$_{2}$, a key parameter that influences the abundance of gas-phase formamide in our model, also requires an experimental measurement. Another possible follow-up study is to determine the spectroscopy of CH$_{3}$OCH$_{2}$, which is an intermediate product in the formation of methyl formate. If this intermediate product will be found observationally, it would confirm that this is the main formation route of methyl formate in the gas phase. Experimental or/and theoretical studies of the processes mentioned above, will be an important step in improving our knowledge of the chemistry of complex organic molecules in prestellar cores.

\acknowledgements
This research has made use of NASA's Astrophysics Data System. AV and PC acknowledge support from the European Research Council (ERC; project PALs 320620). I.J.-S. acknowledges the financial support received from the STFC through an Ernest Rutherford Fellowship (proposal number ST/L004801/1). The authors are thankful to Prof. Eric Herbst for careful reading and valuable suggestions on the paper, to Dr. Patrice Theul\'e, Dr. Jean-Christophe Loison and to Dr. Miwa Goto for the valuable discussions, and to the anonymous referee for the valuable comments that helped to improve the manuscript.

\bibliographystyle{aasjournal}
\bibliography{apj-jour,references}

\clearpage


\begin{table}
  \caption{Important reactions governing chemistry of complex organic molecules in cold environments.}
  \label{tbl:newr}
  \begin{tabular}{l|c|c}\hline\hline
  Reaction & Rate at 10 K, & Reference \\
           & cm$^{3}\cdot$s$^{-1}$ &  \\
  \hline
  OH~+~CH$_{3}$OH~$\rightarrow$~CH$_{3}$O~+~H$_{2}$O                & 1.1(-10) & S13, A16 \\
  OH~+~CH$_{3}$OCH$_{3}$~$\rightarrow$~CH$_{3}$OCH$_{2}$~+~H$_{2}$O & 1.7(-11) & S14 \\
  F~+~CH$_{3}$OCH$_{3}$~$\rightarrow$~CH$_{3}$OCH$_{2}$~+~HF        & 2.0(-10) & HN96 \\
  Cl~+~CH$_{3}$OCH$_{3}$~$\rightarrow$~CH$_{3}$OCH$_{2}$~+~HCl      & 2.0(-10) & W88 \\
  CH$_{3}$OCH$_{2}$~+~O~$\rightarrow$~HCOOCH$_{3}$~+~H              & 2.0(-10) & HN96, S05 \\
  CH$_{3}$O~+~CH$_{3}$~$\rightarrow$~CH$_{3}$OCH$_{3}$~+~h$\nu$     & 3.0(-10) & VH13, B15a \\
  CH~+~CH$_{3}$OH~$\rightarrow$~CH$_{3}$CHO~+~H & 1.8(-8) & J00 \\
  CH~+~CH$_{3}$OH~$\rightarrow$~CH$_{3}$~+~H$_{2}$CO & 1.6(-7) & J00 \\
  C~+~CH$_{3}$CHO~$\rightarrow$~C$_{2}$H$_{4}$~+~CO & 5.4(-10) & HI99 \\
  C~+~H$_{2}$CO~$\rightarrow$~CO~+~CH$_{2}$ & 6.2(-10) & HI99 \\
  CH$_{3}$~+~HCO~$\rightarrow$~CH$_{3}$CHO~+~h$\nu$                 & 5.0(-11) & NIST \\
  NH$_{2}$~+~H$_{2}$CO~$\rightarrow$~NH$_{2}$CHO~+~H                & 2.9(-11) & S17 \\
  \hline
  \end{tabular}

  Notes: a(-b) stands for $a\times10^{-b}$. $\alpha$, $\beta$ and $\gamma$ are the coefficients in the modified Arrhenius expression for the reaction rate coefficient: $k=\alpha\cdot(T/300)^{\beta}\cdot exp(-\gamma/T)$. S13 refers to \citet[][]{Shannon_ea13}, A16 refers to \citet[][]{Antinolo_ea16}, S14 refers to \citet[][]{Shannon_ea14}, HN96 refers to \citet[][]{HoyermannNacke96}, W88 refers to \citet[][]{Wallington_ea88}, S05 refers to \citet[][]{Song_ea05}, VH13 refers to \citet[][]{VasyuninHerbst13_coms}, B15a refers to \citet[][]{Balucani_ea15}, S17 refers to \citet[][]{Skouteris_ea17}, J00 refers to \citet[][]{Johnson_ea00}, HI99 refers to \citet[][]{HusainIoannou91}. URL for NIST Chemical Kinetics Database: http://kinetics.nist.gov/kinetics/index.jsp

\end{table}


\clearpage

\begin{table}
  \caption{Composition of icy mantles averaged over the whole L1544 in Main~model~(MM). Ice constituents with abundances larger than 1\% wrt. solid water are shown}.
  \label{tbl:ice}
  \begin{tabular}{l|c}\hline\hline
             Species & Abundance wrt. \\
                     & solid water, \% \\
    \hline
    H$_{2}$O       & 100 \\
    CO             & 35  \\
    CH$_{3}$OH     & 28  \\
    N$_{2}$        & 26  \\
    H$_{2}$CO      & 19  \\
    NH$_{3}$       & 7   \\
    CH$_{4}$       & 6   \\
    H$_{2}$O$_{2}$ & 4   \\
    \hline
  \end{tabular}
\end{table}

\clearpage

\begin{table}
  \caption{Observed and best-fit modeled fractional abundances of organic species detected in cold interstellar clouds, obtained as ratios of column densities N(X)/N(H$_{2}$). }
  \label{tbl:abu}
  \begin{tabular}{lllll}\hline\hline
             Species &             Main~model~(MM) & \multicolumn{3}{c}{Observations} \\
     &  & L1544 & L1689b & B1-b \\
    \hline
    HCOOCH$_{3}$      & 6.0(-11) & 1.5(-10) J      & 7.4(-10) B  & 2.0(-11) C \\
    CH$_{3}$OCH$_{3}$ & 4.0(-11) & 5.1(-11) J      & 1.3(-10) B  & 2.0(-11) C \\
    NH$_{2}$CHO       & 3.8(-12) & $\le$8.7(-13) J & ---         & ---        \\
    CH$_{3}$CHO       & 5.8(-10) & 2.1(-10) J      & 1.7(-10) B  & 1.0(-11) C \\
    CH$_{3}$O         & 1.0(-10) & 2.7(-11)  J     & 3.3(-11) BF & 4.7(-12) C \\
    HCO               & 4.6(-10) & ---             & 3.6(-10) BF & ---        \\
    H$_{2}$CO         & 4.8(-08) & ---             & 1.3(-09) B  & 4.0(-10) M \\
    CH$_{3}$OH        & 2.7(-08) & 6.0(-9) V       & ---         & 3.1(-09) O \\
    \hline
  \end{tabular}

  Notes:  a(-b) stands for $a\times10^{-b}$. B refers to \citet{Bacmann_ea12}, BF refers to \citet[][]{BacmannFaure16} (the fractional abundances derived as fraction of column densities provided in Table~6 and Table~7 of BF divided by the H$_{2}$ column density value of 3.6$\cdot$10$^{22}$~cm$^{-2}$ taken from \citealt[][]{Bacmann_ea16}), C refers to \citet{Cernicharo_ea12}, J refers to the abundances of COMs measured toward the position of the ``methanol peak'' defined by \citet[][]{JimenezSerra_ea16}, M refers to \citet{Marcelino_ea05}, O refers to \citet{Oeberg_ea10}, and V refers to \citet[][]{Vastel_ea14}. Values for Main~model~(MM) are smoothed for a telescope beam size of 26'' to allow comparison with observed values by \citet[][]{JimenezSerra_ea16}.

\end{table}

\clearpage

\begin{table}
  \caption{Key gas-phase precursors of COMs, and contribution from surface chemistry to their total formation rate at the moment of best agreement at R=0.015~pc in Main~model~(MM)}.
  \label{tbl:surface_contribution}
  \begin{tabular}{l|c|c}\hline\hline
             Species & Surface          & RD efficiency \\
                     & contribution, \% & \\
    \hline
    CH$_{3}$O      & 45  & 3.4(-5) \\
    OH             & 10  & 4.5(-2) \\
    HCO            & 20  & 2.1(-3) \\
    CH$_{3}$       & 3   & 6.2(-1) \\
    C$_{2}$H$_{4}$ & 98  & 6.0(-2) \\
    NH$_{2}$       & 5   & 2.5(-1) \\
    H$_{2}$CO      & 75  & 5.4(-2) \\
    CH$_{3}$OH     & 99  & 6.4(-3) \\
    \hline
  \end{tabular}
\end{table}

\clearpage

\end{document}